\documentclass[aps, prd, floats, floatfix, twocolumn, superscriptaddress, nofootinbib, 10pt]{revtex4-2}

\usepackage[utf8]{inputenc}
\usepackage{amsfonts,amsmath,amssymb,amsthm,mathtools}
\allowdisplaybreaks
\usepackage[dvipsnames,x11names]{xcolor}
\usepackage[sort&compress]{natbib}

\usepackage{graphicx}
\usepackage{makecell}
\usepackage{booktabs}
\usepackage{array}
\usepackage{latexsym}
\usepackage[citecolor=MidnightBlue,linkcolor=MidnightBlue,urlcolor=MidnightBlue,colorlinks=true]{hyperref}
\usepackage[title]{appendix}
\usepackage{aas_macros}
\usepackage{natbib}
\usepackage{tabularx}
\usepackage{xspace}
\usepackage{siunitx}
\usepackage[normalem]{ulem}

\usepackage{afterpage}

\newcommand{\X}{\textcolor{blue}{X}}
	\def\lp{\left(}
	\def\rp{\right)}
	
	\def\abs{\textup{Abs}}
	\def\exp{\textup{exp}}
	
	\def\eos{\mbox{EoS}\xspace}

	\def\Pc{P_c}
	\def\Mb{M_0}
	\def\Rb{R_0}
	
	\def\I{I}
	\def\IS{I_S}
	\def\JS{J_S}
	
	\def\dM{\delta M}
	\def\MS{M_S}
	
	\def\Q{Q}
	\def\QS{Q_S}
	
	\def\L{\lambda_S}
	
	\def\OS{\Omega_S}
	\def\chiS{\chi_S}
	
	\def\Ibar{{\overline{\I}}}
	\def\Qbar{{\overline{\Q}}}
	\def\dMbar{{\overline{\dM}}}
	
	\def\Ibarrot{{\widetilde{\I}}}
	\def\Qbarrot{{\widetilde{\Q}}}
	\def\dMbarrot{{\widetilde{\dM}}}
	
	\def\IbarrotL{{\Ibarrot\lvert_{(\L,\chiS)}}}
	\def\QbarrotL{{\Qbarrot\lvert_{(\L,\chiS)}}}
	\def\IbarrotQbar{{\Ibarrot\lvert_{(\Qbar,\chiS)}}}
	\def\IbarrotQbarrot{{\Ibarrot\lvert_{(\Qbarrot,\chiS)}}}
	
	\def\dMbarrotL{{\dMbarrot\lvert_{(\L,\chiS)}}}
	\def\dMbarrotIbar{{\dMbarrot\lvert_{(\Ibar,\chiS)}}}
	\def\dMbarrotQbar{{\dMbarrot\lvert_{(\Qbar,\chiS)}}}

	\def\IbarL{{\color{\colornorot}{\Ibar\lvert_{(\L)}}}}
	\def\QbarL{{\color{\colornorot}{\Qbar\lvert_{(\L)}}}}
	
	\def\dMbarL{{\color{\colornorot}{\dMbar\lvert_{(\L)}}}}

	\def\yx{$y$-$x$\xspace}
	
	\def\relIL{$I$--Love\xspace}
	\def\relIbarL{$\Ibar$--Love\xspace}
	\def\relIbarrotL{$\Ibarrot$--Love$(\chiS)$\xspace}
	
	\def\relQL{$Q$--Love\xspace}
	\def\relQbarL{$\Qbar$--Love\xspace}
	\def\relQbarrotL{$\Qbarrot$--Love$(\chiS)$\xspace}
	
	\def\relIQ{$I$--$Q$\xspace}
	\def\relIbarQbar{$\Ibar$--$\Qbar$\xspace}
	\def\relIbarrotQbar{$\Ibarrot$--$\Qbar(\chiS)$\xspace}
	\def\relIbarrotQbarrot{$\Ibarrot$--$\Qbarrot(\chiS)$\xspace}
	
	\def\reldML{$\dM$--Love\xspace}
	\def\reldMbarL{$\dMbar$--Love\xspace}
	
	\def\reldMbarrotIbar{$\widetilde{\dM}$--$\overline{I}(\chiS)$\xspace}
	\def\reldMbarrotIbarrot{$\widetilde{\dM}$--$\widetilde{I}(\chiS)$\xspace}
	\def\reldMbarrotQbar{$\widetilde{\dM}$--$\overline{Q}(\chiS)$\xspace}
	\def\reldMbarrotQbarrot{$\widetilde{\dM}$--$\widetilde{Q}(\chiS)$\xspace}

	\def\reldMbarQbar{$\dMbar$--$\Qbar$\xspace}

	\def\reldMbarIbar{$\dMbar$--$\bar{I}$\xspace}
	
	\def\relILQ{$I$--Love--$Q$\xspace}
	\def\relIbarLQbar{$\Ibar$--Love--$\Qbar$\xspace}
	
	\def\relILQdM{$I$--Love--$Q$--$\dM$\xspace}
	\def\relIbarLQbardMbar{$\Ibar$--Love--$\Qbar$--$\dMbar$\xspace}

	\def\rotf{{{\tilde{f}}}}
	
	\def\relQbarrotL{$\Qbarrot$--Love$(\chiS)$\xspace}
	\def\relIbarrotQbarrot{$\Ibarrot$--$\Qbarrot(\chiS)$\xspace}

	\def\relIbarrotLQbarrot{$\Ibarrot$--Love--$\Qbarrot(\chiS)$\xspace}
	\def\any{\omega}
	
	\def\relIbarrotLQbarrotany{$\Ibarrot$--Love--$\Qbarrot(\any)$\xspace}
	
	\def\barred{barred\xspace}
	\def\tilded{tilded\xspace}
	
	\def\colorT{Black}
	\def\colorchakra{Black}
	\def\colornorot{Black}
	
	\def\colorE{Black}
	\def\colorX{Black}
	\def\colorC{Black}
	
	\def\C{{{\color{\colorC}{\textup{std}}}}\xspace}
	
	\def\ISC{{\color{\colorC}{\IS^\C}}}
	\def\QSC{{\color{\colorC}{\QS^\C}}}
	
	\def\IbarLC{{\color{\colorC}{\Ibar\lvert_{(\L)}^\C}}}
	\def\QbarLC{{\color{\colorC}{\Qbar\lvert_{(\L)}^\C}}}

	\def\X{{{\color{\colorX}{\textup{ext}}}}\xspace}
	
	\def\IbarrotX{{\color{\colorX}{\Ibarrot^\X}}}
	\def\QbarrotX{{\color{\colorX}{\Qbarrot^\X}}}
	
	\def\IbarrotLX{{\color{\colorX}{\Ibarrot\lvert_{(\L)}^\X}}}
	\def\QbarrotLX{{\color{\colorX}{\Qbarrot\lvert_{(\L)}^\X}}}
	\def\IbarrotQbarrotX{{\color{\colorX}{\Ibarrot\lvert_{(\Qbarrot,\chiS)}^\X}}}
	
	\def\IbarrotLX{{\color{\colorX}{\Ibarrot\lvert_{(\L,\chiS)}^\X}}}
	\def\QbarrotLX{{\color{\colorX}{\Qbarrot\lvert_{(\L,\chiS)}^\X}}}
	
	\def\T{{{\color{\colorT}{\textup{NP}}}}\xspace}
	
	\def\IbarrotT{{\color{\colorT}{\Ibarrot^{\T}}}}
	\def\QbarrotT{{\color{\colorT}{\Qbarrot^{\T}}}}
	
	\def\IbarrotLT{{\color{\colorT}{\Ibarrot\lvert_{(\L,\chiS)}^\T}}}
	\def\QbarrotLT{{\color{\colorT}{\Qbarrot\lvert_{(\L,\chiS)}^\T}}}
	\def\IbarrotQbarrotT{{\color{\colorT}{\Ibarrot\lvert_{(\Qbarrot,\chiS)}^\T}}}
	\def\IbarrotQbarT{{\color{\colorT}{\Ibarrot\lvert_{(\Qbar,\chiS)}^\T}}}
	\def\IbarrotQbarX{{\color{\colorX}{\Ibarrot\lvert_{(\Qbar,\chiS)}^\X}}}
	
	\def\dMbarrotLT{{\color{\colorT}{\dMbarrot\lvert_{(\L,\chiS)}^\T}}}
	\def\dMbarrotIbarT{{\color{\colorT}{\dMbarrot\lvert_{(\Ibar,\chiS)}^\T}}}
	\def\dMbarrotIbarrotT{{\color{\colorT}{\dMbarrot\lvert_{(\Ibarrot,\chiS)}^\T}}}
	\def\dMbarrotQbarT{{\color{\colorT}{\dMbarrot\lvert_{(\Qbar,\chiS)}^\T}}}
	\def\dMbarrotQbarrotT{{\color{\colorT}{\dMbarrot\lvert_{(\Qbarrot,\chiS)}^\T}}}
	
	\def\fILrotT{{\color{\colorT}{\mathcal{B}}}}
	\def\fQLrotT{{\color{\colorT}{\mathcal{C}}}}
	\def\fIQrotT{{\color{\colorT}{\mathcal{D}}}}
	\def\fIQT{{\color{\colorT}{\mathcal{F}}}}
	
	\def\fdMLT{{\color{\colorT}{\mathcal{K}}}}
	\def\fdMIT{{\color{\colorT}{\mathcal{L}}}}
	\def\fdMIrotT{{\color{\colorT}{\mathcal{M}}}}
	\def\fdMQT{{\color{\colorT}{\mathcal{N}}}}
	\def\fdMQrotT{{\color{\colorT}{\mathcal{P}}}}
	
	\def\IST{{\color{\colorT}{\IS^{\T}}}}
	\def\QST{{\color{\colorT}{\QS^{\T}}}}

	\def\chakra{{{\color{\colorchakra}{\textup{RNS}}}}\xspace}

	\def\IbarrotchakraQbarrot{{\color{\colorchakra}{\Ibarrot}\lvert^{\chakra}_{(\Qbarrot,\chiS)}}}
	
	\def\pol{{\color{\colornorot}{\mathrm{Pol}}}}
	\def\polyx{{\color{\colornorot}{\pol_{y(x)}}}}
	\def\yx{{\color{\colornorot}{y\lvert_{(x)}}}}

	\def\polIbarL{{\color{\colornorot}{\pol_{\Ibar(\L)}}}}
	\def\polQbarL{{\color{\colornorot}{\pol_{\Qbar(\L)}}}}
	\def\polIbarQbar{{\color{\colornorot}{\pol_{\Ibar(\Qbar)}}}}
	
	\def\poldMbarL{{\color{\colornorot}{\pol_{\dMbar(\L)}}}}
	\def\poldMbarIbar{{\color{\colornorot}{\pol_{\dMbar(\Ibar)}}}}
	\def\poldMbarQbar{{\color{\colornorot}{\pol_{\dMbar(\Qbar)}}}}
	
	\def\ayx{{\color{\colornorot}{a_{y(x)}}}}
	\def\byx{{\color{\colornorot}{b_{y(x)}}}}
	\def\cyx{{\color{\colornorot}{c_{y(x)}}}}
	\def\dyx{{\color{\colornorot}{d_{y(x)}}}}
	\def\eyx{{\color{\colornorot}{e_{y(x)}}}}
	
	\def\aIbarL{{\color{\colornorot}{a_{\Ibar(\L)}}}}
	\def\bIbarL{{\color{\colornorot}{b_{\Ibar(\L)}}}}
	\def\cIbarL{{\color{\colornorot}{c_{\Ibar(\L)}}}}
	\def\dIbarL{{\color{\colornorot}{d_{\Ibar(\L)}}}}
	\def\eIbarL{{\color{\colornorot}{e_{\Ibar(\L)}}}}
	
	\def\adMbarL{{\color{\colornorot}{a_{\dMbar(\L)}}}}
	\def\bdMbarL{{\color{\colornorot}{b_{\dMbar(\L)}}}}
	\def\cdMbarL{{\color{\colornorot}{c_{\dMbar(\L)}}}}
	\def\ddMbarL{{\color{\colornorot}{d_{\dMbar(\L)}}}}
	\def\edMbarL{{\color{\colornorot}{e_{\dMbar(\L)}}}}

	\def\error{\mathcal{E}}

	\def\errorIbarrotLTX{\error_{\IS(\L,\chiS)}^{\T/\X}}
	\def\errorQbarrotLTX{\error_{\QS(\L,\chiS)}^{\T/\X}}
	
	\def\errorIbarrotLTX{\error_{\Ibarrot(\L,\chiS)}^{\T/\X}}
	\def\errorQbarrotLTX{\error_{\Qbarrot(\L,\chiS)}^{\T/\X}}
	\def\errorIbarrotQbarTX{\error_{\Ibarrot(\Qbar,\chiS)}^{\T/\X}}

	\def\errorIbarrotQbarrotchakra{\error_{\Ibarrot(\Qbarrot,\chiS)}^{\T/\chakra}}

	\def\E{{\color{\colorE}{\textup{full}}}\xspace}
	
	\def\rotfE{{\color{\colorE}{\rotf^\E{}}}}
	\def\MSE{{\color{\colorE}{\MS^\E{}}}}
	\def\ISE{{\color{\colorE}{\IS^\E{}}}}
	\def\QSE{{\color{\colorE}{\QS^\E{}}}}
	\def\JSE{{\color{\colorE}{\JS^\E{}}}}
	
	\def\chiSE{{\color{\colorE}{\chiS^\E{}}}}
	
	\def\IbarrotE{{\color{\colorE}{\Ibarrot{}^\E}}}
	\def\QbarrotE{{\color{\colorE}{\Qbarrot{}^\E}}}
	
	\def\relIbarrotLE{{\color{\colorE}{$\Ibarrot^\E$--Love$^\E(\chiS^\E)$}}\xspace}
	\def\relQbarrotLE{{\color{\colorE}{$\Qbarrot^\E$--Love$^\E(\chiS^\E)$}}\xspace}
	\def\relIbarrotQbarrotE{{\color{\colorE}{$\Ibarrot^\E$--$\Qbarrot^\E(\chiS^\E)$}}\xspace}

	\def\IbarrotEchakraQbarrot{{\color{\colorchakra}{\IbarrotE{}\lvert^\chakra_{(\QbarrotE,\chiSE)}}}}
	
	\def\MbX{{\color{\colorX}{\Mb^\X}}}
	\def\ISX{{\color{\colorX}{\IS^\X}}}
	\def\QSX{{\color{\colorX}{\QS^\X}}}

	\def\IbarrotX{{\color{\colorX}{\Ibarrot^\X}}}
	\def\QbarrotX{{\color{\colorX}{\Qbarrot^\X}}}
	
	\newcolumntype{C}{>{\centering\arraybackslash}m{6em}}
	
	\newcounter{mnotecount}%
	\newcommand{\mnote}[1]%
	{\protect{\stepcounter{mnotecount}}$^{\mbox{\footnotesize $\bullet$\themnotecount}}$ 
		\marginpar{%
			\raggedright\tiny\textit{
				$\!\!\!\!\!\!\,\bullet$\themnotecount: #1} }}
	
	\makeatletter
	\newcommand*{\textvcenter}{\@ifstar{\@tempswatrue\text@vcenter}{\@tempswafalse\text@vcenter}}
	\newcommand*{\text@vcenter}[1]{\mbox{$\m@th\vcenter{\setbox\z@=\hbox{#1}\if@tempswa\dp\z@\z@\fi\box\z@}$}}
	
	\newcommand*{\showmathaxis}{%
		\setbox\z@=\hbox{$a$}%
		\@tempdima\fontdimen22\textfont2\relax
		\@tempdimb\@tempdima
		\advance\@tempdima.2\p@
		\advance\@tempdimb-.2\p@
		\leavevmode\rlap{\vrule height\@tempdima depth-\@tempdimb width10cm}%
	}
	\makeatother

\usepackage{tikz}
\usetikzlibrary{shapes.geometric, arrows}

\begin{document}

\title{Rotation-dependent \texorpdfstring{\relILQdM}{I-Love-Q-dM} relations in perturbation theory}

\author{Eneko Aranguren}%
\email{eneko.aranguren@ehu.eus}%
\affiliation{Department of Physics, University of the Basque Country UPV/EHU, P.O.~Box 644, 48080 Bilbao, Basque Country, Spain}%
\affiliation{EHU Quantum Center, University of the Basque Country UPV/EHU, P.O.~Box 644, 48080 Bilbao, Basque Country, Spain}%

\begin{abstract}
	The so-called \relILQ relations link some normalized versions of the
	moment of inertia, the Love number, and the quadrupole moment of a star.
	These relations, in principle, enable the inference of two of the quantities given the third.
	However, their use has been limited because the normalized versions of the multipole moments
	rely on the static mass derived from the Tolman--Oppenheimer--Volkoff equation, which is not directly observable.
	In this work, using perturbation theory, we find that the \relILQ relations can also be formulated in terms of an alternative set of normalized quantities
	that do not depend on the static mass, but on the actual (observable) mass.
\end{abstract}
\maketitle

\section{Introduction}

In the early XX century, the mathematician Augustus E. H. Love provided a perturbative description of the Earth's yielding to the tidal forces exerted by the Sun and the Moon~\cite{love_1909}.
In this seminal work, he introduced what we now call the surficial and gravitational Love numbers, $h$ and $k$, respectively.
The former characterizes the vertical deformation of the Earth's surface under the effect of a tidal field,
while the latter is related to the change in the gravitational potential.
The geophysics investigations carried out by Love were later extended by T. Shida in~\cite{toshi_1912},
who stressed the need to consider a third tidal quantity $l$
describing the horizontal displacement of the tidal response.
The tidal perturbations can be expressed as a sum of Legendre polynomials,
being
$h_\ell$, $k_\ell$ and $l_\ell$ the corresponding components,
that provide a complete description of the Earth's linear deformation response to the tides.

This geophysical analysis was later extended to describe the tidal deformation of a stellar binary system
through the apsidal motion constants~\cite{brooker_1955_newtonian_love,sterne_1939,kopal_1978,motz_1953}, that have a direct correspondence with the Love numbers $k_\ell$ ~\cite{Csizmadia_2019_apsidal_love}.
The tidal deformability $\L$, a quantity related to the leading-order Love number $k_2$,
was found in~\cite{Flanagan_Hinderer_2008} to have a potentially measurable impact on the gravitational wave signal emitted during
the late
inspiral of a stellar binary.
The Love numbers known so far, however, were restricted to the limitations of Newtonian physics.
The prospect of observational detection thus motivated the search for a relativistic formulation of tidal deformations,
a description that was pursued in~\cite{damour_nagar_2009} and \cite{poisson_tidal_2009}.

In General Relativity there are no known exact analytical solutions of isolated, realistic rotating stars,
	so the study of such configurations must be done using perturbation theory.
The perturbative expansion to second order in the angular velocity $\OS$~\cite{hartlethorne, reina:2014_amending} is commonly referred to as the Hartle--Thorne model.
Solving the field equations for a given Equation of State (\eos) and central pressure $\Pc$,
we can calculate the different multipolar properties at each order,
such as the Tolman--Oppenheimer--Volkoff (TOV) mass $\Mb(\Pc)$~\cite{tolman,oppenheimer_volkoff} of the background (static) 
configuration,
the moment of inertia $\IS(\Pc)$,
the quadrupole moment $\QS(\Pc,\OS)$,
and the mass of the rotating star $\MS(\Pc,\OS)$.
The TOV mass $\Mb(\Pc)$ is a mathematical artifact, with no observational impact whatsoever,
and is related to the actual mass $\MS(\Pc,\OS)$ by a factor $\OS^2\,\dM(\Pc)$.

In~\cite{yagi_yunes_iloveq} it was found that some 
normalized counterparts of 
$\IS(\Pc)$, $\L(\Pc)$, and $\QS(\Pc,\OS)$ of compact stars were related in an \eos-insensitive way, through some relations called \relILQ.
It was claimed therein that these relations could be used to infer two of the quantities involved in the \relILQ trio, out of the third one alone.
These normalized quantities 
depend on the TOV mass $\Mb(\Pc)$,
so to translate from the normalized quantities to the actual multipolar moments, the value of $\Mb(\Pc)$ needs to be computed first.
However, the fact that $\Mb(\Pc)$ is not directly observable hinders the applicability of the relations,
an obstacle that is usually overcome in the literature by assuming that $\Mb(\Pc) = \MS(\Pc,\OS)$ (see e.g.~\cite{yagi_kyutoku_2014}),
which is only true in the static limit $\OS = 0$.
This approach (that we will call `standard') has been justified in~\cite{yagi_yunes:prd88,pani_berti:scalartensor},
on the grounds that, up to the time of their publication, the observed binaries had a slow rotation,
with spin parameters $\chiS<0.1$.
Nevertheless, recent detections of gravitational wave signals have identified millisecond pulsars whose
spin can attain values as high as $\chiS\sim0.4$ (see~\cite{Hessels}).

In~\cite{reina:dM} a new universal relation was found between some normalized version of $\dM(\Pc)$ and the Love number.
It was mentioned therein that this new \reldML relation could enable the inference of the mass $\Mb(\Pc)$
needed for a correct application of the \relILQ relations
--- we refer to this approach as the `extended approach'.
This claim was later developed in~\cite{Aranguren:dM},
where a comparison between the standard and extended approaches was carried out.
It was shown that the multipolar quantities derived within the extended approach were more accurate than
those obtained from the standard counterpart,
with the same conclusion holding for the inference of the \eos parameters.
This improvement in precision is achieved
purely through a different implementation of the calculations
--- without having to enhance the sensitivity of the gravitational wave detectors.

In~\cite{Chakrabarti:2013,Pappas:2013naa} a rotation-dependent \relIQ relation computed within a fully numerical approach was proposed.
In this numerical framework, the multipolar properties of the star are computed to full extent,
without resorting to perturbation theory, 
as opposed to the approach adopted in the original relations from~\cite{yagi_yunes_iloveq}.
We say this \relIQ relation is rotation-dependent because the normalized quantities
involved depend on (the full, nonperturbative version of) $\MS(\Pc,\OS)$ instead of $\Mb(\Pc)$.
This fully numerical rotation-dependent relation dodges the need to compute $\Mb(\Pc)$ in the first place.
Nevertheless, since the moment of inertia cannot be detected from gravitational wave observations~\cite{yagi:2014_tests},
and the quadrupole moment has a small effect on the waveform~\cite{Chatziioannou:2018},
the usefulness of the above \relIQ relation gets tainted.
Therefore, it is usually more convenient to work with \relIL and \relQL,
as the tidal deformability gets inprinted in the gravitational wave signal.

In the present work, we find a complete set of rotation-dependent \relILQdM relations,
in analogy with the \relIQ couple found in~\cite{Chakrabarti:2013,Pappas:2013naa},
but computed through perturbation theory to second order.
We do this in two different ways:
first, we propose a set of rotation-dependent relations purely from the already known \relILQdM quartet,
and then we second (and extend) these results
numerically using different \eos's and $\Pc$'s.

The article is arranged as follows:
In Sec.~\ref{sec:moments}, we review the multipolar moments of rotating stars to second order in perturbation theory,
which constitute the building blocks of the universal relations.
In Sec.~\ref{sec:universal}, we describe the standard approach to apply the universal relations,
and the associated difficulties that motivated the construction of the (nonperturbative) rotation-dependent \relIQ couple in~\cite{Chakrabarti:2013,Pappas:2013naa}.
Based on the universal relations for $\dM$ found in~\cite{reina:dM, Aranguren:dM},
in Sec.~\ref{sec:extended} we elaborate on the application of the extended approach
considering separately $\OS$ and $\chiS$ as the known rotation magnitudes.
Within this framework, we then derive some rotation-dependent relations in terms of $\chiS$.
Section~\ref{sec:rotation} focuses on the numerical computation of the rotation-dependent relations using perturbation theory
and compares the results with those derived from the extended \relILQdM set, and with a second order truncation of the fully numerical \relIQ relation provided in~\cite{Chakrabarti:2013}.
In Sec.~\ref{sec:rotation_dependent_dM} we elaborate on the rotation-dependent relations involving $\dM$.
Finally, Sec.~\ref{sec:conclusions} will close the work with a discussion of the main results.

\textcolor{black}{For a smoother reading, we use geometric units ($G = c = 1)$,
	and move all the tables to Appendix~\ref{app:tables}, where
a glossary of the main terms and symbols is also provided
for quick reference.}

\section{Multipolar properties of the star}\label{sec:moments}
The study of slowly rotating isolated stars is usually tackled within perturbation theory to second order in the angular velocity
(see \cite{aranguren:2022,Reina:2015_homogeneous,berti_2005}).
This perturbative approach corresponds to the Hartle--Thorne model~\cite{hartlethorne} with the correction in~\cite{reina:2014_amending},
a formalism that was later derived from first principles in~\cite{MRV1,MRV2}.

In this setting, 
	first and second order stationary and axisymmetric perturbation tensors
	are constructed over a static and spherically symmetric background configuration.
	The gauge freedom present in the perturbative framework is then conveniently chosen to
	find the field equations and the matching conditions between the inner and outer regions of the star.
The system of equations is then closed provided a perturbed \eos is specified with a unique barotropic relation
governing the dependence between the energy density $E$ and pressure $P$ at every order.
Finally, for a given \eos describing the background configuration, the entire solution is uniquely determined for some value of the central pressure $\Pc$ and angular velocity $\OS$.

The interior problem in the background configuration (order zero) is described by the so-called
Tolman--Oppenheimer--Volkoff (TOV) equation~\cite{tolman,oppenheimer_volkoff}.
Provided the star has a finite radius,
which is not a priori ensured~\cite{Rendall_1991},
the mass $\Mb(\Pc)$ and radius $\Rb(\Pc)$
are fixed by imposing the matching conditions with a Schwarzschild exterior.

Likewise, the first order interior problem is matched with the first order vacuum exterior
to give the angular momentum $\JS(\Pc,\OS)$, which is related with the moment of inertia via
\begin{align}\label{IS}
    \IS(\Pc) = \frac{\JS(\Pc,\OS)}{\OS}.
\end{align}

The second order problem can be expanded into the Legendre polynomials with only the $l=0$ and $l=2$ terms.
The matching conditions in the $l=0$ problem fix the value of $\dM(\Pc)$ (see the amendment in~\cite{reina:2014_amending}),
which is the contribution of the mass at second order,
so that the total mass is
\begin{align}\label{MS}
    \MS(\Pc,\OS) = \Mb(\Pc) + \OS^2\,\dM(\Pc).
\end{align}
Equivalently, the $l=2$ sector determines the value of $\Q(\Pc)$, 
related to the quadrupole moment via
\begin{align}\label{QS}
    \QS(\Pc,\OS) = \OS^2\,Q(\Pc),
\end{align}
which measures the deformation of the gravitational field~\cite{hartlethorne}.

Even if the Hartle--Thorne model serves to describe isolated and rotating stars to second order,
the perturbative scheme can be reproduced analogously for static configurations embedded in a tidal field
~\cite{Flanagan_Hinderer_2008,Hinderer_2008,damour_nagar_2009,poisson_tidal_2009,Hinderer_2009}.
Depending on the parity of the perturbations under an inversion of the spatial coordinates $\vec{x}\rightarrow-\vec{x}$,
these are classified as electric (even --- do not change), and magnetic (odd --- do change).
For the purposes of the present work, we focus on the even-parity sector at first order in the perturbative expansion,
which is characterized by the (leading-order) quadrupolar Love number $k_2(\Pc)$,
that quantifies how the shape of a spherically symmetric gravitational field is affected by the presence of a companion star.
We introduce also the dimensionless tidal deformability $\L(\Pc)$, defined by
\begin{align}\label{L}
    \L(\Pc) := \frac{2\Rb(\Pc)^5}{3\Mb(\Pc)^5}k_2(\Pc),
\end{align}
which plays a key role in gravitational-wave astronomy.

Now, let us consider the exact multipole moments of the star
	$\JSE$, $\QSE$, $\MSE$,
	which are the generalizations of $\JS$, $\QS$ and $\MS$ to the full, nonperturbative solution of the field equations.
	We assume that near $\OS=0$ these exact quantities can be expanded as follows
	\begin{alignat}{2}
		&\JSE(\Pc,\OS) &&= \OS\,\IS(\Pc) + \OS^3\,\IS^{(3)}(\Pc) + O(\OS^5),\label{JSE}\\
		&\QSE(\Pc,\OS) &&= \OS^2\,\Q(\Pc) + \OS^4\,\Q^{(4)}(\Pc) + O(\OS^6),\label{QSE}\\
		&\MSE(\Pc,\OS) &&= \Mb(\Pc) + \OS^2\,\dM(\Pc) + O(\OS^4),\label{MSE}
	\end{alignat}
	where the superscript $(i)$ denotes that the corresponding term enters at order $i$ in the perturbative expansion,
	while those terms with no indices correspond to the magnitudes~\eqref{IS}-\eqref{QS} considered so far.
	The dimensionless spin parameter $\chiSE(\Pc,\OS)$, defined as
	\begin{align}
		\chiSE(\Pc,\OS) := \frac{\JSE(\Pc,\OS)}{\MSE(\Pc,\OS)^2},\label{chiSE_not_expanded}
	\end{align}
	can also be expanded perturbatively via
	\begin{alignat}{2}
		\chiSE(\Pc,\OS) &= \chiS(\Pc,\OS) \notag\\
		&+ \OS^3\lp\frac{\IS^{(3)}(\Pc)}{\IS(\Pc)}-2\frac{\dM(\Pc)}{\Mb(\Pc)}\rp\,\frac{\IS(\Pc)}{\Mb(\Pc)^2}\notag\\
		&+ O(\OS^5),\label{chiSE}
	\end{alignat}
	where
	\begin{align}
		\chiS(\Pc,\OS) := \frac{\IS(\Pc)\OS}{\Mb(\Pc)^2},\label{chiS}
	\end{align}
	is the second order truncation of $\chiSE(\Pc,\OS)$.

\section{Universal relations and the nonperturbative rotation-dependent set}\label{sec:universal}
A set of `universal' (or \eos-insensitive) relations involving the moment of inertia $\IS(\Pc)$,
the tidal deformability $\L$ and the quadrupole moment $\QS$ were found in \cite{yagi_yunes_iloveq}.
These relations are called the \relILQ relations,
and relate not the bare quantities~\eqref{IS}-\eqref{QS}, but the dimensionless and rotation-independent counterparts
(that we call `\barred' quantities)
\begin{align}
    &\Ibar(\Pc):=\frac{\JS(\Pc,\OS)}{\OS\,\Mb(\Pc)^3}=\frac{\IS(\Pc)}{\Mb(\Pc)^3},\label{Ibar}\\
    &\Qbar(\Pc):=\frac{\QS(\Pc,\OS)\,\Mb(\Pc)}{\JS(\Pc,\OS)^2}=\frac{\Q(\Pc)\,\Mb(\Pc)}{\IS(\Pc)^2},\label{Qbar}
\end{align}
together with $\L$, which plays the role of `Love'.
The \relILQ relations work for quark stars and neutron stars with cold \eos (see e.g.,~\cite{Guedes_2024}),
but do not hold in general for other stellar configurations,
such as non-corrotating double fluid superfluid neutron stars~\cite{Yeung:2021wvt} (amended in~\cite{aranguren:2022}),
or strongly magnetized stars~\cite{haskell}.
The universal relations are described by relations of the form
\begin{align}
    \ln\yx = \polyx,\label{lnyx}
\end{align}
where
\begin{align}
	\polyx &:= \ayx + \byx \ln x + \cyx \lp\ln x\rp^2 \notag\\
	&+ \dyx \lp\ln x\rp^3 + \eyx \lp\ln x\rp^4,\label{polyx}
\end{align}
for any $x,y\in\{\Ibar,\L,\Qbar\}$.
The polynomial constants $\{\ayx,\byx,\cyx,\dyx,$ $\eyx\}$ for some combinations of $x$ and $y$ are given in Table~\ref{tab:polyx}.
From now on we will use the notation \relIbarLQbar to refer to the \relILQ relations.
The usefulness will become more evident later.

The \relIbarLQbar relations enable the inference of two quantities out of a third one alone.
Among the three quantities involved in the \relIbarLQbar trio, the tidal deformability is the one that can be more easily obtained,
since the weighted average tidal deformability $\L^\textup{eff}$~\cite{Flanagan_Hinderer_2008,Favata_2014} of a binary star, given by
\begin{align}
    \frac{13}{16}(M_A + M_B)^5 \L^{\textup{eff}} &= (M_A + 12 M_B)M_A^4 \lambda_{S_A}\notag\\
    &+ (12 M_A + M_B)M_B^4 \lambda_{S_B},\label{L_effective}
\end{align}
is inprinted in the gravitational wave signal during the inspiral~\cite{abbott:2017,Zhu:2018,GW170817:measurements, GW190425}.
One can then use Bayesian analysis to extract the individual tidal deformabilities of the constituent
stars\footnote{A similar procedure is followed to obtain the individual masses and rotation magnitudes from some effective mass (called the chirp mass) $\mathcal{M}$ \cite{finn:chirp_mass,damour:chirp_mass,cutler:chirp_mass,poisson:chirp_mass}, and spin parameter $\chiS^\textup{eff}$ \cite{damour:chis_eff,ajith:chis_eff,santamaria:chis_eff}.
}.
Therefore, we define the triplet
\begin{align}
    \{\L,\MS,\chiS\},\label{inputset}
\end{align}
as the input set for the entire work,
and focus, in particular, on the extraction of $\IS$ and $\QS$ derived from there.

Once $\Ibar$ and $\Qbar$ have been inferred from the \relIbarL and \relQbarL relations,
the corresponding values of $\IS$ and $\QS$
are computed through the definitions~\eqref{Ibar}-\eqref{Qbar},
\begin{align}
    &\IS = \Mb^3\, \Ibar,\label{IS_from_Ibar_theory}\\
    &\QS = \OS^2\,\Mb^5\, \Ibar^2\, \Qbar,\label{QS_from_Qbar_OS_theory}
\end{align}
or equivalently,
\begin{align}
    \QS = \chiS^2\, \Mb^3\, \Qbar.\label{QS_from_Qbar_chiS_theory}
\end{align}
These expressions depend on the TOV mass $\Mb(\Pc)$, which
is not a physical property of the star, and differs from the actual (observable) mass $\MS(\Pc,\OS)$ via~\eqref{MS}.
Therefore, in order to translate from the \barred $\Ibar$, $\Qbar$ to the actual $\IS$, $\QS$,
we need to obtain the value of $\Mb(\Pc)$ first.

\subsection{Standard approach}\label{subsec:standard}
The standard approach used in the literature --- that we denote with an `\C' ---
overcomes this difficulty by assuming that $\Mb(\Pc)=\MS(\Pc,\OS)$ in~\eqref{Ibar}-\eqref{Qbar} (see, for example,~\cite{yagi_yunes:prd88}).
The procedure reads as follows.
First, using the Love number from the input set~\eqref{inputset},
one resorts to the \relIbarL and \relQbarL relations to obtain $\Ibar$ and $\Qbar$ via
\begin{align}
	&\ln\IbarLC = \polIbarL,\label{standard_IbarLC}\\
	&\ln\QbarLC = \polQbarL.
\end{align}
Then, replacing $\Mb(\Pc)$ with $\MS(\Pc,\OS)$ in~\eqref{IS_from_Ibar_theory}-\eqref{QS_from_Qbar_OS_theory},
one obtains
\begin{align}
	& \ISC = \MS^3\,\IbarLC,\label{ISC}\\
	& \QSC = \OS^2\,\MS^5\,(\IbarLC)^2\,\QbarLC.\label{QSC}
\end{align}
These are the values of $\IS$ and $\QS$ obtained within the standard approach.
Analogously, one can express $\QSC$ in terms of $\chiS$ using again the substitution $\Mb(\Pc)\rightarrow\MS(\Pc,\OS)$
in~\eqref{QS_from_Qbar_chiS_theory} to find
\begin{align}
	\QSC = \chiS^2\, \MS^3\, \QbarLC.\label{QS_from_Qbar_chiS_standard}
\end{align}
This approach was acknowledged to break the universality of the \relIbarLQbar relations in~\cite{Doneva:2013rha},
and the quantitative deviations were studied in detail in~\cite{Aranguren:dM}.

\subsection{Rotation-dependent relations in the fully numerical setup}
If we want to use $\MS(\Pc,\OS)$ instead of $\Mb(\Pc)$ in the universal relations,
we first need to construct the analogs of $\Ibar$ and $\Qbar$, with $\Mb(\Pc)$ replaced by $\MS(\Pc,\OS)$.
These new quantities, that we denote with a tilde, are given by
\begin{align}
	&\Ibarrot(\Pc,\OS) :=\frac{\JS(\Pc,\OS)}{\OS\,\MS(\Pc,\OS)^3}= \frac{\IS(\Pc)}{\MS(\Pc,\OS)^3},\label{Ibarrot}\\
	&\Qbarrot(\Pc,\OS) := \frac{\QS(\Pc,\OS)\, \MS(\Pc,\OS)}{\JS(\Pc,\OS)^2} \notag\\
	&\qquad\qquad\,\,\,\,= \frac{\Q(\Pc)\MS(\Pc,\OS)}{\IS(\Pc)^2}.\label{Qbarrot}
\end{align}
Then, we need to see whether these tilded quantities, which depend on the rotation, can be related in an \eos-insensitive way.
That is, we need to see if some rotation-dependent \relIbarrotLQbarrotany relations exist, in terms of some rotation-parameter $\any$.

This path was actually explored in~\cite{Chakrabarti:2013,Pappas:2013naa}
using the fully numerical (nonperturbative) results from the RNS code~\cite{RNS} (an approach that we denote with the superscript `RNS').
It was found therein that a universal relation emerges between some normalized versions of $\JSE$ and $\QSE$,
given by
\textcolor{black}{\begin{align}
	&\IbarrotE(\Pc,\OS):= \frac{\JSE(\Pc,\OS)}{\OS \MSE(\Pc,\OS)^3},\label{IbarrotE_notexpanded}\\
	&\QbarrotE(\Pc,\OS) := \frac{\QSE(\Pc,\OS) \MSE(\Pc,\OS)}{\JSE(\Pc,\OS)^2},\label{QbarrotE_notexpanded}
\end{align}}
in terms of the spin parameter $\chiSE$~\eqref{chiSE_not_expanded} via
\begin{align}
	\ln\IbarrotEchakraQbarrot\,:=\,\sum_{i=0}^4\sum_{j=0}^4\mathcal{A}_{ij}\lp\chiSE\rp^i \lp\ln\QbarrotE\rp^j,\label{lnIbarrotEchakra}
\end{align}
where $\mathcal{A}$ is a constant matrix whose elements, as computed in~\cite{Chakrabarti:2013}\footnote{\textcolor{black}{The rotation-dependent \relIbarrotQbarrotE relation from~\cite{Pappas:2013naa}
		could also be considered,
		as it was found therein to be compatible with the one provided in \cite{Chakrabarti:2013}.}},
are given in Table~\ref{tab:polyxrotchakra}.
In~\cite{Chakrabarti:2013} an alternative relation was also found for some dimensionless frequency $\rotfE$.
However, it was acknowledged that using $\chiSE$ as the rotation magnitude works best for the universality,
so we stick to that choice as well.
The remaining relations \relIbarrotLE and \relQbarrotLE,
where `Love$^\E$' refers to the fully numerical counterpart of `Love',
are precisely those required given the input set~\eqref{inputset},
but have not been obtained in the literature so far.

For later reference,
we use the expressions~\eqref{JSE}-\eqref{MSE} to expand the tilded quantities~\eqref{IbarrotE_notexpanded}-\eqref{QbarrotE_notexpanded} in $\OS$.
This yields
\begin{align}
	&\IbarrotE =\Ibar+\OS^2\lp\frac{\IS^{(3)}}{\IS}-3\frac{\dM}{\Mb}\rp\,\Ibar + O(\OS^4),\label{IbarrotE}\\
	&\QbarrotE = \Qbar + \OS^2\lp\frac{\Q^{(4)}}{\Q} -2\frac{\IS^{(3)}}{\IS}+\frac{\dM}{\Mb}\rp\Qbar +O(\OS^4).\label{QbarrotE}
\end{align}

\subsection{Implications of the standard approach}\label{subsec:implications_standard}
We briefly return to the discussion in Sec.~\ref{subsec:standard}
	to provide an alternative interpretation of the standard approach
	in terms of $\Ibarrot$~\eqref{Ibarrot} and $\Qbarrot$~\eqref{Qbarrot}.
	These tilded quantities are precisely $\Ibar$~\eqref{Ibar} and $\Qbar$~\eqref{Qbar}
	with the approximation $\Mb(\Pc)\rightarrow\MS(\Pc,\OS)$, that was employed in the standard approach.
	Therefore, we can regard the standard approach as a hybrid construction that mixes the rotation-dependent $\Ibarrot$, $\Qbarrot$
	with the rotation-independent relations~\eqref{lnyx}-\eqref{polyx}, whose polynomial coefficients $\{\ayx$, $\byx$, $\cyx$, $\dyx$, $\eyx\}$ were computed for $\Ibar$ and $\Qbar$.
	
The tilded quantities admit an expansion in $\OS$ that is given by
\begin{align}
	&\Ibarrot = \Ibar-\OS^2\,\frac{3\dM}{\Mb}\,\Ibar + O(\OS^4),\label{Ibarrot_expanded}\\
	&\Qbarrot = \Qbar+\OS^2\,\frac{\dM}{\Mb}\,\Qbar,\label{Qbarrot_expanded}
\end{align}
so we expect that the relative errors of the normalized quantities within the standard approach grow with $\OS^2$ via
\begin{align}
	&\frac{\Ibar-\Ibarrot}{\Ibar}\approx\OS^2\,\frac{3\dM}{\Mb},\label{relative_error_I}\\
	&\frac{\Qbar-\Qbarrot}{\Qbar}=-\OS^2\,\frac{\dM}{\Mb}.\label{relative_error_Q}
\end{align}

\section{Extended approach}\label{sec:extended}
A major step was given in the implementation of the \relIbarLQbar relations
with the discovery of a new universal relation between some dimensionless and rotation-independent counterpart of $\dM$,
given by
\begin{align}
    \dMbar:=\frac{\dM \Mb^3}{\IS^2},\label{dMbar}
\end{align}
and $\L$ in~\cite{reina:dM}.
This \reldMbarL relation was later complemented with the remaining \reldMbarIbar and \reldMbarQbar relations computed in~\cite{Aranguren:dM}.
We include the corresponding polynomial coefficients in Table~\ref{tab:polyx}.

It was claimed in~\cite{reina:dM} that the \reldMbarL relation found therein could be used to infer the value of $\Mb$
needed in the normalizations~\eqref{Ibar}-\eqref{Qbar}.
The step-by-step implementation of this
`extended approach' (denoted with an `\X')
was developed in~\cite{Aranguren:dM},
and the procedure reads as follows.

First, using the input set~\eqref{inputset}, we resort to the \reldMbarL relation
\begin{align}
    \ln\dMbarL &= \adMbarL + \bdMbarL \ln\L\notag\\
    &+ \cdMbarL \lp\ln\L\rp^2 + \ddMbarL \lp\ln\L\rp^3\notag\\
    &+ \edMbarL \lp\ln\L\rp^4,\label{lndMbarL}
\end{align}
to infer the value of $\dMbar$,
which can be written in terms of $\OS$ via [see~\eqref{dMbar}]
\begin{align}
	\dMbar=\frac{\MS-\Mb}{\OS^2 \Ibar^2\Mb^3}.\label{dMbar_OS}
\end{align}
This equation depends on $\Ibar$, that we obtain from the \relIbarL relation
\begin{align}
	\ln\IbarL &= \aIbarL + \bIbarL \ln\L\notag\\
	&+ \cIbarL \lp\ln\L\rp^2 + \dIbarL \lp\ln\L\rp^3\notag\\
	&+ \eIbarL \lp\ln\L\rp^4.\label{lnIL}
\end{align}
Once $\dMbar$ and $\Ibar$ are known,
we can solve~\eqref{dMbar_OS} for $\Mb$ to obtain the following (depressed) cubic equation
\begin{align}
    \Mb^3 + \frac{1}{\dMbar\OS^2 \Ibar^2} \Mb - \frac{\MS}{\dMbar\OS^2 \Ibar^2} = 0.\label{Mb_cubic}
\end{align}
We follow the notation in \cite{Garrett_Birkhoff} to calculate the discriminant $D$ of the above equation
\begin{align}\label{discriminant}
    D = - \frac{4}{\dMbar^3 \OS^6 \Ibar^6}-\frac{27 \MS^2}{\dMbar^2 \OS^4 \Ibar^4},
\end{align}
which is always negative provided\footnote{For a stellar configuration with $E + P > 0$, the mass $\Mb$ is always positive.
In analogy to the Newtonian approach, the parameter $\dM$ represents the mass that needs to be injected into the star so that the central pressure is unchanged with the deformation~\cite{chandrasekhar_1933_dM,chandrasekhar_1962_dM}.
It is then motivated on physical grounds that $\dM$ has to be positive, although it has not been proven yet.
Nevertheless, $\dM$ is indeed positive in all cases considered.}
$\dMbar>0$.
Since $D<0$, the cubic equation~\eqref{Mb_cubic} has two complex conjugate roots,
so $\Mb$ is given by the remaining real root --- which we denote by $\MbX$.
Lastly, the corresponding multipolar quantities
in the extended approach are computed through
\begin{align}
    &\ISX =  \lp\MbX\rp^3\,\Ibar,\label{IS_extended}\\
    &\QSX = \OS^2 \lp\MbX\rp^5 \Ibar^2\,\Qbar,\label{QS_extended_OS}
\end{align}
where 
$\Qbar$ is obtained from the 
\relQbarL relation.

Note that within the above procedure, the computation of $\MbX$~\eqref{Mb_cubic} involves the \relIbarL relation.
This is also the case for $\QSX$ from~\eqref{QS_extended_OS},
where the \relIbarL relation enters both via $\MbX$ and through $\Ibar$ explicitly.
The fact that we need not only the \reldMbarL relation but also \relIbarL to obtain $\MbX$ increases the errors.
This, however, can be improved by repeating the previous calculations using the dimensionless spin parameter
$\chiS$~\eqref{chiS} instead of $\OS$,
as done in~\cite{Aranguren:2025_proceedings}.
Indeed, using $\chiS$, the definition~\eqref{dMbar} reads
\begin{align}
    \dMbar=\frac{\MS-\Mb}{\chiS^2 \Mb},\label{dMbar_chiS}
\end{align}
which has the unique solution
\begin{align}
    \MbX = \frac{\MS}{\dMbar\chiS^2 + 1}.\label{Mb_chiS}
\end{align}
We will use this expression for $\MbX$ from now on,
so we will no longer resort to the solution of~\eqref{Mb_cubic} to define $\MbX$.
Once $\Ibar$ and $\Qbar$ are known from the \relIbarL and \relQbarL relations,
the actual moment of inertia is obtained via~\eqref{IS_extended},
while for the quadrupole moment we have
\begin{align}\label{QS_extended_chiS}
    \QSX = \chiS^2 \, \lp\MbX\rp^3 \Qbar.
\end{align}
Note that now $\Ibar$ does not enter~\eqref{Mb_chiS} nor \eqref{QS_extended_chiS},
so the accuracy is expected to be higher, as claimed in~\cite{Aranguren:2025_proceedings}.

\subsection{Rotation-dependent relations from the \texorpdfstring{\relIbarLQbardMbar}{I-Love-Q-dM} set}\label{rotation_dependent_extended}
Let us now use the extended \relIbarLQbardMbar set 
to compute the rotation-dependent \relIbarrotLQbarrot relations.
For that, we consider the following algebraic identities
\begin{align}
    &\ln\Ibarrot = \ln\Ibar - 3\ln\lp1+\chiS^2\dMbar\rp,\label{lnIbarrot}\\
    &\ln\Qbarrot = \ln\Qbar + \ln\lp1+\chiS^2\dMbar\rp,\label{lnQbarrot}
\end{align}
that follow from the definitions of $\Ibarrot$~\eqref{Ibarrot} and $\Qbarrot$~\eqref{Qbarrot}.
For any $x,y\in\{\Ibarrot,\L,\Qbarrot,\Qbar\}$,
we use the notation $\ln y\lvert_{(x,\chiS)}$
to remark that we are considering the rotation-dependent relation in terms of $\chiS$.
From~\eqref{lnIbarrot}-\eqref{lnQbarrot} the \relIbarrotL and \relQbarrotL relations in the extended approach read
\begin{alignat}{2}
    &\ln\IbarrotLX &&= \ln\IbarL - 3\ln\lp1+\chiS^2\,\dMbarL\rp,\label{lnIbarrotX_L}\\
    &\ln\QbarrotLX &&= \ln\QbarL + \ln\lp1+\chiS^2\,\dMbarL\rp,\label{lnQbarrotX_L}
\end{alignat}
which, after using~\eqref{lnyx}, can be expressed as
\begin{alignat}{2}
    &\ln\IbarrotLX &&= \polIbarL\label{lnIbarrotX_L_pol} \\
    & &&- 3\ln\lp1+\chiS^2 \,\exp\lp\poldMbarL\rp\rp,\notag\\
    &\ln\QbarrotLX &&= \polQbarL \label{lnQbarrotX_L_pol}\\
    & &&+ \ln\lp1+\chiS^2 \,\exp\lp\poldMbarL\rp\rp.\notag
\end{alignat}
Equivalently, we can construct the \relIbarrotQbar relation from~\eqref{lnIbarrot}
\begin{align}
    \ln\IbarrotQbarX &= \polIbarQbar\label{lnIbarrotX_Qbar}\\
    &- 3\ln\lp1+\chiS^2\,\exp\lp\poldMbarQbar\rp\rp.\notag
\end{align}
Observe that the \relIbarrotQbarrot relation
cannot be derived analytically within the extended approach.
This is because the $\Ibar$ and $\dMbar$ that appear in the expansion of $\Ibarrot$~\eqref{lnIbarrot}
must be inferred using the \relIbarLQbardMbar relations,
for which a \barred quantity is required as the input.
In fact, this limitation is present in the computation of any \tilded--\tilded relation,
such as \reldMbarrotIbarrot and \reldMbarrotQbarrot
(we will explore the rotation-dependent relations involving $\dM$ in Sec.~\ref{sec:rotation_dependent_dM}).
Therefore, the extended approach only allows for the computation of \tilded--\barred relations.

Finally, using the input set~\eqref{inputset},
the multipolar quantities are recovered through
\begin{align}
    &\ISX = \MS^3\,\IbarrotX,\label{ISX}\\
    &\QSX = \chiS^2\,\frac{(\MbX)^4}{\MS}\,\QbarrotX.\label{QSX}
\end{align}

\section{Numerically computed rotation-dependent relations}\label{sec:rotation}
In this section, we investigate computationally the unexplored rotation-dependent \relIbarrotLQbarrot relations using the perturbative setup.

Since we are considering $\chiS$ as the rotation magnitude, 
the argument $\chiS$ in $\ln y\lvert_{(x,\chiS)}$ might feel redundant.
However, as happened in \cite{Chakrabarti:2013}, additional rotation-dependent relations may be found in terms of different rotation parameters.
Therefore, we prefer to make the dependence on $\chiS$ explicit at all times.

Let us first consider a representative set of \eos's.
We study single fluid neutron star configurations with the \eos's A~\cite{pandharipande:1971A} and FPS~\cite{Lorenz_EOS}.
To describe quark stars, we take the Colpi \& Miller \eos from \cite{Colpi:1992},
together with MB1, MB2 and MB3, which correspond to the SQM1-3 models in~\cite{Lattimer_2001} after using the linear \eos described in~\cite{Zdunik:2000}.
Finally, double-fluid superfluid neutron stars are represented by NL3~\cite{Fattoyev:2010}, GM1~\cite{Glendenning:1991}, the polytropic \eos Poly from~\cite{comer1999}, and the toy model TOY suggested in~\cite{aranguren:2022}.

For each \eos, and for different values of $\Pc$ and $\chiS$,
we compute $\Ibarrot$, $\L$, $\Qbarrot$, and $\Qbar$.
Then, we pack the pieces of data into sets:
$\{\ln\Ibarrot,\ln\L,\chiS\}$,
$\{\ln\Qbarrot,\ln\L,\chiS\}$,
$\{\ln\Ibarrot,\ln\Qbarrot,\chiS\}$,
and $\{\ln\Ibarrot,\ln\Qbar,\chiS\}$,
and interpolate numerically to construct the functions
$\ln\IbarrotL$,
$\ln\QbarrotL$,
$\ln\IbarrotQbarrot$
and $\ln\IbarrotQbar$, respectively.
The corresponding rotation-dependent relations are then approximated as fourth order polynomial functions via
\begin{alignat}{2}
    &\ln\IbarrotLT&&=\,\sum_{i=0}^4\sum_{j=0}^4\fILrotT_{ij}\,\chiS^i\,\lp\ln\L\rp^j,\label{lnIbarrotTL}\\
    &\ln\QbarrotLT&&=\,\sum_{i=0}^4\sum_{j=0}^4\fQLrotT_{ij}\,\chiS^i\,\lp\ln\L\rp^j,\label{lnQbarrotTL}\\
    &\ln\IbarrotQbarrotT&&=\,\sum_{i=0}^4\sum_{j=0}^4\fIQrotT_{ij}\,\chiS^i\,\lp\ln\Qbarrot\rp^j,\label{lnIbarrotQbarrotT}\\
    &\ln\IbarrotQbarT&&=\,\sum_{i=0}^4\sum_{j=0}^4\fIQT_{ij}\,\chiS^i\,\lp\ln\Qbar\rp^j,\label{lnIbarrotQbarT}
\end{alignat}
where the coefficients $\{\fILrotT_{ij},\fQLrotT_{ij},\fIQrotT_{ij},\fIQT_{ij}\}$ are given in Table~\ref{tab:truncated}.
This approach, where the rotation-dependent relations are computed numerically using perturbation theory (with the \eos's considered above),
is referred to as the `numerical perturbative approach', denoted with the superscript `\T'.

\begin{figure*}[htb!]
    \centering
    \includegraphics[width=0.85\textwidth]{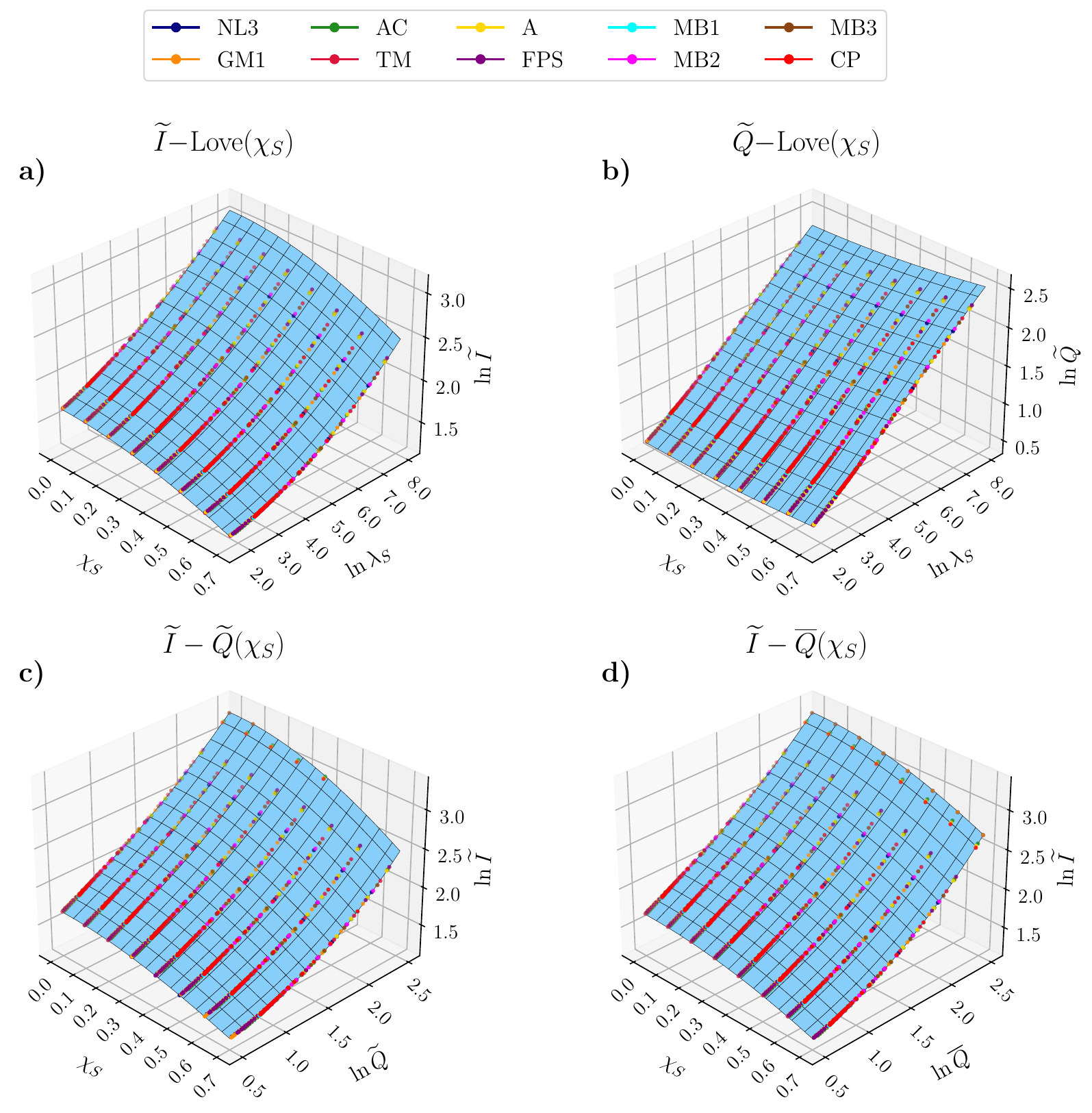}
    \caption{Different relations of the \relIbarrotLQbarrot set.
             The dots in each plot represent the results from the individual \eos's,
             while the surfaces are the polynomial functions from~\eqref{lnIbarrotTL}-\eqref{lnIbarrotQbarT}.
             Panel a) shows $\ln\IbarrotLT$~\eqref{lnIbarrotTL}, b) $\ln\QbarrotLT$~\eqref{lnQbarrotTL}, c) $\ln\IbarrotQbarrotT$~\eqref{lnIbarrotQbarrotT},
             and d) represents $\ln\IbarrotQbarT$~\eqref{lnIbarrotQbarT}.}
    \label{fig:rotation_dependent_relations}
\end{figure*}

In Fig.~\ref{fig:rotation_dependent_relations} we show the numerical results from the \eos's together with
the fourth order polynomial surfaces from~\eqref{lnIbarrotTL}-\eqref{lnIbarrotQbarT}.
The results show that there exist some universal relations between
$\Ibarrot$, $\L$, $\Qbarrot$ and $\Qbar$,
with a dependence on the spin parameter $\chiS$.

The multipolar moments are recovered using
\begin{alignat}{2}
    &\IST &&= \MS^3\,\IbarrotT,\label{IST}\\
    &\QST &&= \chiS^2 \,\frac{\lp\Mb\rp^4}{\MS}\, \QbarrotT.\label{QST_chiS}
\end{alignat}
Observe that due to the dependence of $\QST$ on $\Mb$,
the numerical perturbative approach cannot be used to determine $\QST$ on its own.
This was also the case for the $\QSX$ in~\eqref{QSX}.
However, the extended approach can infer the value of $\MbX$ through~\eqref{Mb_chiS},
so the inference of $\QS$ is possible in that case.

\begin{figure*}[htb!]
    \centering
    \includegraphics[width=\textwidth]{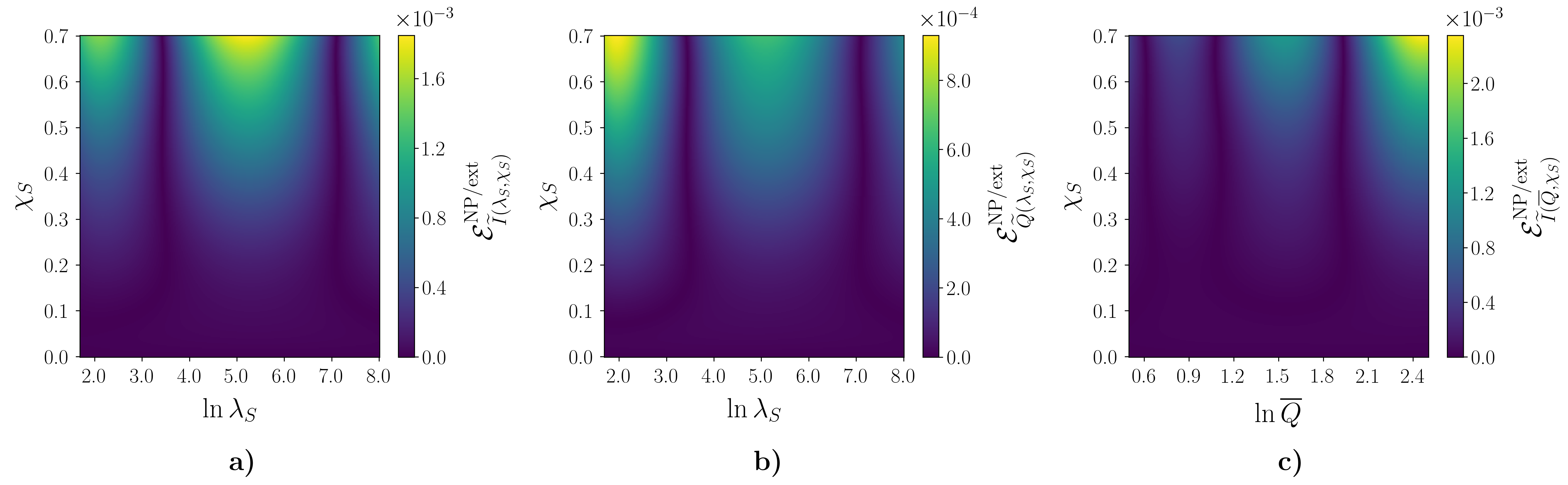}
    \caption{Panel a) represents $\errorIbarrotLTX$~\eqref{errorILTX}, b) $\errorQbarrotLTX$~\eqref{errorQLTX},
    and c) $\errorIbarrotQbarTX$~\eqref{errorIQTX}.
    The color bars show that the relative errors are below $\sim0.2\%$ in all cases.}
    \label{fig:H_A}
\end{figure*}

To compare the rotation-dependent universal relations computed within the extended and the numerical perturbative approaches, we use the following relative errors
\begin{align}
    &\errorIbarrotLTX:=\abs\lp\frac{\ln\IbarrotLT-\ln\IbarrotLX}{\ln\IbarrotLT}\rp,\label{errorILTX}\\
    &\errorQbarrotLTX:=\abs\lp\frac{\ln\QbarrotLT-\ln\QbarrotLX}{\ln\QbarrotLT}\rp,\label{errorQLTX}\\
    &\errorIbarrotQbarTX:=\abs\lp\frac{\ln\IbarrotQbarT-\ln\IbarrotQbarX}{\ln\IbarrotQbarT}\rp,\label{errorIQTX}
\end{align}
which we illustrate in Fig.~\ref{fig:H_A}.
Note that the extended approach does not provide $\ln\IbarrotQbarrotX$,
so~\eqref{lnIbarrotQbarrotT} lacks an extended counterpart to compare with.
The results show that the analytical and numerical approaches have relative errors smaller than $\sim0.2\%$ in all cases.

\subsection{Comparing the \relIbarrotQbarrot relation from the numerical perturbative approach with the fully numerical result}
We now want to compare the relation $\ln\IbarrotQbarrotT$~\eqref{lnIbarrotQbarrotT},
with $\ln\IbarrotEchakraQbarrot$~\eqref{lnIbarrotEchakra}.
This comparison, however, requires some preliminary steps.
The relation $\ln\IbarrotQbarrotT$ depends on $\Ibarrot$, $\Qbarrot$ and $\chiS$,
which are quantities obtained within second order perturbation theory,
whereas $\ln\IbarrotEchakraQbarrot$ is expressed in terms of $\IbarrotE$, $\QbarrotE$ and $\chiSE$,
which correspond to the fully numerical counterparts of $\Ibarrot$, $\Qbarrot$ and $\chiS$, respectively [see~\eqref{IbarrotE}-\eqref{QbarrotE} and ~\eqref{chiSE}].
In order to compare analogous quantities, we consider instead a truncated version of $\ln\IbarrotEchakraQbarrot$.
For that, we first express $\Ibar$ and $\Qbar$ in terms of $\Ibarrot$ and $\Qbarrot$ using~\eqref{Ibarrot_expanded}-\eqref{Qbarrot_expanded}
\begin{align}
	&\Ibar = \Ibarrot \lp1+\OS^2\frac{3\dM}{\Mb}\rp + O(\OS^4),\\
	&\Qbar = \Qbarrot \lp1-\OS^2\frac{\dM}{\Mb}\rp + O(\OS^4).
\end{align}
Then, we substitute this into $\IbarrotE$~\eqref{IbarrotE} and $\QbarrotE$~\eqref{QbarrotE} to get
\begin{align}
	&\IbarrotE = \Ibarrot\lp1+\OS^2\frac{\IS^{(3)}}{\IS}\rp + O(\OS^4),\\
	&\QbarrotE = \Qbarrot\lp1+\OS^2\lp\frac{\Q^{(4)}}{\Q}-2\frac{\IS^{(3)}}{\IS}\rp\rp + O(\OS^4).
\end{align}
Replacing these expressions, together with $\chiSE$~\eqref{chiSE}, into $\ln\IbarrotEchakraQbarrot$~\eqref{lnIbarrotEchakra}, we obtain
\begin{align}
    &\ln\Ibarrot = \sum_{i=0}^{4}\sum_{j=0}^{4}\mathcal{A}_{ij}\,\chiS^i\,\lp\ln\Qbarrot\rp^j\notag\\
    &-\lp\frac{\IS^{(3)}}{\IS}+\lp2\frac{\IS^{(3)}}{\IS}-\frac{\Q^{(4)}}{\Q}\rp\sum_{i=1}^{4}i\,\mathcal{A}_{0i}\lp\ln\Qbar\rp^{i-1}\rp\OS^2\notag\\
    &+ O(\OS^4).\label{lnIE_expansion}
\end{align}
Finally, we keep only the terms that involve quantities up to second order ---
omitting those terms with $\IS^{(3)}$ and $\QS^{(4)}$.
That is, we define
\begin{align}
    \ln\IbarrotchakraQbarrot := \sum_{i=0}^4 \sum_{j=0}^4 \mathcal{A}_{ij}\,\chiS^i\,\lp\ln\Qbarrot\rp^j,\label{lnIbarrotchakra}
\end{align}
as the second order truncation of $\ln\IbarrotEchakraQbarrot$~\eqref{lnIbarrotEchakra}.
The difference between the full expansion~\eqref{lnIE_expansion} and the truncation~\eqref{lnIbarrotchakra} is precisely
\begin{align}
    &-\lp\frac{\IS^{(3)}}{\IS}+\lp2\frac{\IS^{(3)}}{\IS}-\frac{\Q^{(4)}}{\Q}\rp\sum_{i=1}^{4}i\,\mathcal{A}_{0i}\lp\ln\Qbar\rp^{i-1}\rp\OS^2\notag\\
    &+ O(\OS^4),
\end{align}
which corresponds to the error that arises from truncating $\ln\IbarrotEchakraQbarrot$ to
$\ln\IbarrotchakraQbarrot$.
Since~\eqref{lnIbarrotchakra} still maintains the original coefficients $\mathcal{A}_{ij}$
--- that were constructed for the fully numerical quantities\footnote{This was also the idea behind the standard approach,
	where the coefficients of the \relIbarLQbar relations were employed to infer $\Ibarrot$ and $\Qbarrot$ instead of $\Ibar$ and $\Qbar$.} $\IbarrotE$ and $\QbarrotE$ ---
the results from $\ln\IbarrotQbarrotT$ and $\ln\IbarrotchakraQbarrot$
are expected not to coincide.
Nevertheless, this comparison can still shed light on how accurately
the solution from perturbation theory to second order approximates the fully numerical counterpart.

\begin{figure*}[htb!]
    \centering
    \includegraphics[width=0.8\linewidth]{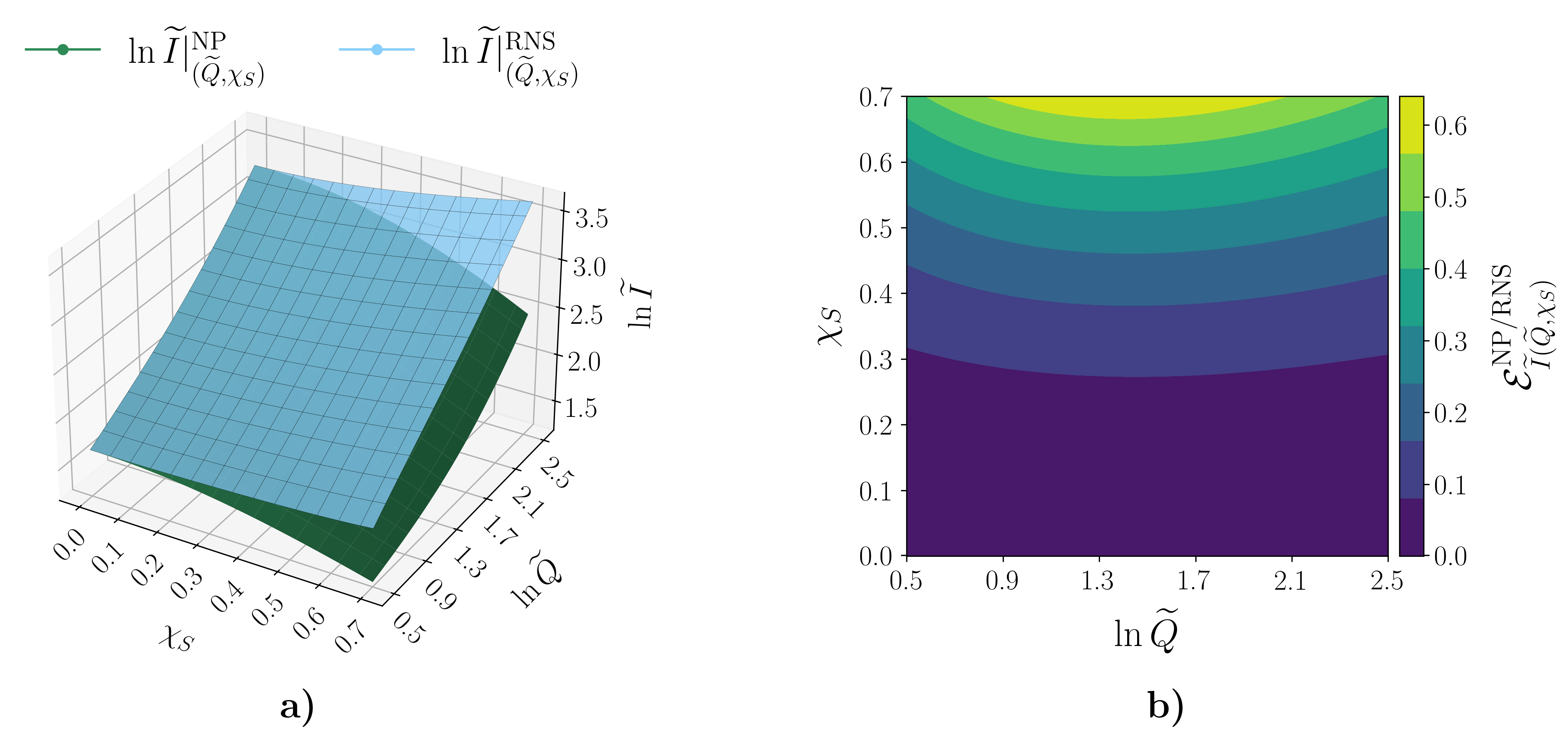}
    \caption{Panel a) represents $\ln\IbarrotQbarrotT$~\eqref{lnIbarrotQbarrotT} (green surface) and $\ln\IbarrotchakraQbarrot$~\eqref{lnIbarrotchakra} (blue surface),
    which is the restriction of $\ln\IbarrotEchakraQbarrot$ to second order in $\OS$.
    Panel b) shows the relative error between both surfaces as defined in~\eqref{errorIQchakra}.}
    \label{fig:chakra}
\end{figure*}

The relative error between $\ln\IbarrotQbarrotT$ and $\ln\IbarrotchakraQbarrot$, defined as
\begin{align}
    \errorIbarrotQbarrotchakra:=\abs\lp\frac{\ln\IbarrotQbarrotT-\ln\IbarrotchakraQbarrot}{\ln\IbarrotQbarrotT}\rp,\label{errorIQchakra}
\end{align}
is shown in Fig.~\ref{fig:chakra}.
One can see that the relative difference between the second order truncation of the fully numerical \relIbarrotQbarrotE relation
and the perturbative counterpart is small ($\lesssim0.1$) for $\chiS\lesssim0.3$,
and increases gradually with $\chiS$,
reaching values of up to $\sim0.6$ at $\chiS=0.7$.

\section{The rotation-dependent relations for \texorpdfstring{$\dMbarrot$}{dM}}\label{sec:rotation_dependent_dM}
\begin{figure*}[htb!]
    \centering
    \includegraphics[width=0.85\textwidth]{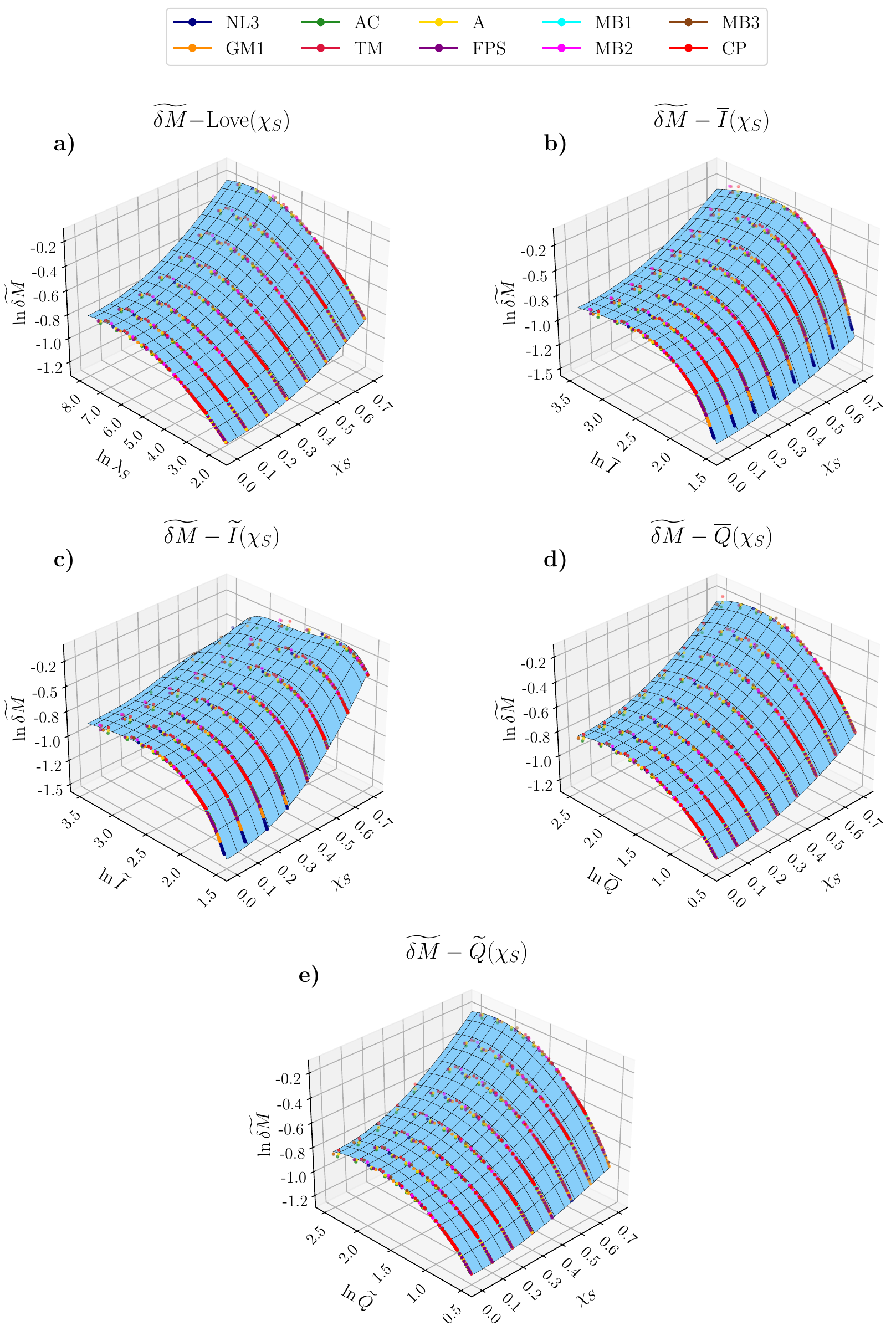}
    \caption{All possible relations involving the quantity $\dMbarrot$.
             The dots in each plot represent the results from all the \eos's,
             while the surfaces are the polynomial functions from~\eqref{lndMbarrotLT}-\eqref{lndMbarrotQbarrotT}.
             Panel a) shows $\ln\dMbarrotLT$~\eqref{lndMbarrotLT}, b) $\ln\dMbarrotIbarT$~\eqref{lndMbarrotIbarT}, c) $\ln\dMbarrotIbarrotT$~\eqref{lndMbarrotIbarrotT},
             d) $\ln\dMbarrotQbarT$~\eqref{lndMbarrotQbarT}, and e) represents $\ln\dMbarrotQbarrotT$~\eqref{lndMbarrotQbarrotT}.}
    \label{fig:rotation_dependent_relations_dM}
\end{figure*}

The parameter $\dM$ can also be normalized with $\MS$ instead of $\Mb$
\begin{align}
    \dMbarrot:=\frac{\dM\MS^3}{\IS^2}.\label{dMbarrot}
\end{align}
This way, we find
\begin{align}
    \ln\dMbarrot = \ln\dMbar + 3\ln\lp1+\chiS^2\,\dMbar\rp,
\end{align}
that from~\eqref{polyx} can be expressed as
\begin{alignat}{2}
    &\ln\dMbarrotL &&= \poldMbarL\label{dMbarrotL}\\
    & &&+3\ln\lp1+\chiS^2\,\exp\lp\poldMbarL\rp\rp,\notag\\
    &\ln\dMbarrotIbar &&= \poldMbarIbar\label{dMbarrotIbar}\\
    & &&+3\ln\lp1+\chiS^2\,\exp\lp\poldMbarIbar\rp\rp,\notag\\
    &\ln\dMbarrotQbar &&= \poldMbarQbar \label{dMbarrotQbar}\\
    & &&+3\ln\lp1+\chiS^2\,\exp\lp\poldMbarQbar\rp\rp.\notag
\end{alignat}
These analytical results follow from the extended approach developed in Sec.~\ref{rotation_dependent_extended}.
The existence of rotation-dependent relations for $\dMbarrot$ can also be checked numerically, as was done with the \relIbarrotLQbarrot relations, using the numerical perturbative approach in Sec.~\ref{sec:rotation}.
Repeating the steps described therein,
we obtain
\begin{alignat}{2}
    &\ln\dMbarrotLT       &&= \sum_{i=0}^4\sum_{j=0}^4 \fdMLT_{ij}   \,\chiS^i\,\lp\ln\L\rp^j,      \label{lndMbarrotLT}\\
    &\ln\dMbarrotIbarT    &&= \sum_{i=0}^4\sum_{j=0}^4 \fdMIT_{ij}   \,\chiS^i\,\lp\ln\Ibar\rp^j,   \label{lndMbarrotIbarT}\\
    &\ln\dMbarrotIbarrotT &&= \sum_{i=0}^4\sum_{j=0}^4 \fdMIrotT_{ij}\,\chiS^i\,\lp\ln\Ibarrot\rp^j,\label{lndMbarrotIbarrotT}\\
    &\ln\dMbarrotQbarT    &&= \sum_{i=0}^4\sum_{j=0}^4 \fdMQT_{ij}   \,\chiS^i\,\lp\ln\Qbar\rp^j,   \label{lndMbarrotQbarT}\\
    &\ln\dMbarrotQbarrotT &&= \sum_{i=0}^4\sum_{j=0}^4 \fdMQrotT_{ij}\,\chiS^i\,\lp\ln\Qbarrot\rp^j,\label{lndMbarrotQbarrotT}
\end{alignat}
where the coefficients ${\fdMLT_{ij}, \fdMIT_{ij}, \fdMIrotT_{ij}, \fdMQT_{ij}, \fdMQrotT_{ij}}$ are given\footnote{Since the universality of the \reldMbarIbar and \reldMbarQbar
	is significantly weaker for high values of $\Ibar$ and $\Qbar$,
	in~\cite{Aranguren:dM} the coefficients of \reldMbarIbar and \reldMbarQbar
	were computed considering only the stellar configurations for which $\Ibar\leq4$ and $\Qbar\leq3$, respectively.
	To have compatible results in the $\chiS=0$ limit, we follow the same criterion here,
	and therefore fit only the data with $\Ibar\leq4$, $\Ibarrot\leq4$, $\Qbar\leq3$ and $\Qbarrot\leq3$
	in \reldMbarrotIbar, \reldMbarrotIbarrot, \reldMbarrotQbar and \reldMbarrotQbarrot, respectively.}
	in Table~\ref{tab:truncated_dM}.
	In Fig.~\ref{fig:rotation_dependent_relations_dM} we show the polynomial surfaces~\eqref{lndMbarrotLT}-\eqref{lndMbarrotQbarrotT}, together with the numerical results from the \eos's.

Using the input set~\eqref{inputset}
one can infer $\dMbarrotL$ through either~\eqref{dMbarrotL} or \eqref{lndMbarrotLT}.
Then, from the definition~\eqref{dMbarrot}, one finds
\begin{align}
    \frac{\dMbarrot\chiS^2}{\MS^3}\,\Mb^4 + \Mb - \MS = 0,
\end{align}
that may serve to obtain the value of $\Mb$.
However, we do not intend to explore these relations any further,
since their usefulness is restricted to the derivation of $\Mb$,
which has already been addressed within the extended approach.

\section{Summary and Conclusions}\label{sec:conclusions}
The main result of the present work is the introduction of the rotation-dependent \relIbarrotLQbarrot relations
computed to second order in perturbation theory.

To begin with, we have defined the \tilded quantities $\Ibarrot$~\eqref{Ibarrot}, $\Qbarrot$~\eqref{Qbarrot}
which are analogous to the \barred quantities $\Ibar$~\eqref{Ibar}, $\Qbar$~\eqref{Qbar} 
with $\Mb(\Pc)$ replaced by $\MS(\Pc,\OS)$.
Then, we have found that some universal relations involving the \tilded quantities exist
as functions of the spin parameter $\chiS$.
This has been done in two different ways.

On the one hand,
we have shown that $\Ibarrot$ and $\Qbarrot$ can be expressed in terms of $\Ibar$, $\Qbar$, $\chiS$ and $\dMbar$,
thus proving the existence of the \relIbarrotL, \relQbarrotL and \relIbarrotQbar relations using solely the already known \relIbarLQbardMbar set.
We call this the extended approach.

On the other hand, we have employed a large number of \eos's and ${\Pc}$'s to verify numerically that
the quantities $\Ibarrot$, $\Qbarrot$, Love, and $\Qbar$ from perturbation theory
can be grouped into pairs forming \eos-insensitive functions of $\chiS$ (see Fig.~\ref{fig:rotation_dependent_relations}),
thereby corroborating
the existence of \relIbarrotL, \relQbarrotL, and \relIbarrotQbar,
as well as the \relIbarrotQbarrot relation, that was not provided in the extended approach.
We call this the numerical perturbative approach,
as the results have been computed numerically using perturbation theory to second order.

As a final step, we have compared the results derived from the extended and the numerical perturbative approaches
and found that they are compatible,
with relative errors of order $\lesssim0.2\%$ (see Fig.~\ref{fig:H_A}).

A rotation-dependent \relIbarrotQbarrotE relation was already found within a fully numerical (nonperturbative) approach in~\cite{Chakrabarti:2013,Pappas:2013naa}.
Since the \relIbarrotQbarrot relation computed in the present work is perturbative,
to provide a meaningful comparison,
we have truncated the \relIbarrotQbarrotE relation from~\cite{Chakrabarti:2013} to second order in $\OS$.
This direct, albeit naive, comparison has been shown in Fig.~\ref{fig:chakra}.
As expected, these relations are not compatible for high values of $\chiS$,
since the results computed within perturbation theory depart from the fully numerical counterpart for rapidly rotating configurations.
However, there is very good agreement for $\chiS\lesssim0.3$.

Lastly, in Sec.~\ref{sec:rotation_dependent_dM} we have shown that \eos-insensitive relations also exist for $\dMbarrot$~\eqref{dMbarrot} [the rotation-dependent counterpart of $\dMbar$~\eqref{dMbar}],
opening the way to alternative derivations of the TOV mass $\Mb$.

The numerical analysis developed in~\cite{Aranguren:dM,Aranguren:2025_proceedings}
found that the approximation $\Mb(\Pc) = \MS(\Pc,\OS)$ (that we call the standard approach) in the definition of the normalized quantities
introduces errors in the inference of the multipolar moments which are significantly larger than those obtained from 
the extended approach.
In particular, comparing the results shown in Fig.~1 in~\cite{Aranguren:2025_proceedings},
we see that the relative errors of the inferred $\IS$ and $\QS$ 
with the actual values computed from perturbation theory are up to $\sim25$ times higher in the standard approach than in the 
extended counterpart.
Regarding the inference of the \eos, it was shown in~\cite{Aranguren:dM} that the relative errors in the standard approach can be up to $\sim77$ higher than in the extended approach (see Sec.~V.~A. therein).

Since the numerically computed rotation-dependent relations suggested here are compatible with the results from the extended approach (see Fig.~\ref{fig:H_A}),
the conclusions from~\cite{Aranguren:dM,Aranguren:2025_proceedings} ---
that motivated the replacement of the standard approach in favor of the extended counterpart ---
can also be translated to the present rotation-dependent \relIbarrotLQbarrot set~\eqref{lnIbarrotTL}-\eqref{lnIbarrotQbarT},
which may in fact be useful in future gravitational-wave analyses.

The universal \relILQdM relations were computed to second order in perturbation theory,
so their validity can only be considered within the validity of the perturbative framework.
In this regard,
it was concluded in~\cite{berti:validity_HT} that the second order truncation of the exact theory is very reliable
for most astrophysical applications.
Also, it is usually mentioned that the `slow rotation' regime
is delimited by the condition $\OS<<\Omega_\textup{K}$ (see e.g.~\cite{posada:2023bnm}),
where $\Omega_\textup{K}$ is the Kepler (mass-shedding) limit.
The recent work~\cite{Kwon:2025:validity_HT}, however, suggests that for various stellar properties,
perturbation theory to second order in $\OS$ is not sufficiently accurate
even in the slow-rotation regime.
In any case, from the results presented in~\cite{Aranguren:dM} [and from the errors~\eqref{relative_error_I}-\eqref{relative_error_Q}],
it is clear that the extended and the numerical perturbative approaches yield more accurate results than the standard counterpart,
also when the use of perturbation theory is valid.

Additional universal relations are also known to exist for higher order multipoles~\cite{yagi_kyutoku_2014}.
Therefore, the results presented here may encourage the search for higher order (or even fully numerical) counterparts of the second order rotation-dependent \relIbarrotL and \relQbarrotL universal relations.

\section{Acknowledgements}
I am grateful to Raül Vera for the careful reading of the manuscript, and for the useful comments and suggestions.
I would also like to thank José A. Font, Nicolas Sanchis-Gual and Daniela Doneva for useful interactions.
I acknowledge financial support from the Basque Government Grant No. PRE\_2024\_2\_0078.
Work supported by the Basque Government (Grant No. IT1628-22) and
Spanish Agencia Estatal de Investigación (Grant No. PID2021-123226NB-I00 funded by “ERDF A way of making Europe” and MCIN/AEI/10.13039/501100011033).

\appendix
\section{Tables and glossary}\label{app:tables}
All the supporting tables mentioned in the main work have been collected here for convenience.
\textcolor{black}{We also provide a glossary of the relevant terms and notations.}

\begin{table*}[htb!]
	\renewcommand{\arraystretch}{1.5}
	\centering\label{key}
	\setlength{\tabcolsep}{15pt}
	\begin{tabular*}{\linewidth}{@{\extracolsep{\fill}}c
			S[table-format=1.2e1]
			S[table-format=1.2e1]
			S[table-format=1.2e1]
			S[table-format=1.2e1]
			S[table-format=1.2e1]}
		\hline\hline
		& \multicolumn{1}{c}{$j=0$} & \multicolumn{1}{c}{$j=1$} & \multicolumn{1}{c}{$j=2$} & \multicolumn{1}{c}{$j=3$} & \multicolumn{1}{c}{$j=4$} \\
		\hline
		$\mathcal{A}_{0j}$ & 1.35   & 0.697   & -0.143  & 0.0994  & -0.0124 \\
		$\mathcal{A}_{1j}$ & 0.3541 & -1.435  & 1.721   & -0.8199 & 0.1348 \\
		$\mathcal{A}_{2j}$ & -1.871 & 8.385   & -9.343  & 4.429   & -0.7355 \\
		$\mathcal{A}_{3j}$ & 3.034  & -14.75  & 18.14   & -8.782  & 1.460 \\
		$\mathcal{A}_{4j}$ & -1.860 & 10.05   & -12.65  & 6.100   & -1.008 \\
		\hline\hline
	\end{tabular*}
	\caption{The coefficients $\mathcal{A}_{ij}$~\eqref{lnIbarrotchakra} given in~\cite{Chakrabarti:2013}.}
	\label{tab:polyxrotchakra}
\end{table*}

\begin{table*}[htb!]
	\setlength{\tabcolsep}{15pt}
	\renewcommand{\arraystretch}{1.5}
	\centering
	\begin{tabular*}{\linewidth}{@{\extracolsep{\fill}} c c S[table-format=1.2e1] S[table-format=1.2e1] S[table-format=1.2e1] S[table-format=1.2e1] S[table-format=1.2e1]}
		\hline\hline
		$y$    & $x$     & $\ayx$ & $\byx$ & $\cyx$  & $\dyx$  & $\eyx$ \\
		\hline
		$\Ibar$  & $\L$   & 1.47   & 0.0817 & 0.0149  & 2.87e-4  & -3.64e-5 \\
		$\Qbar$  & $\L$   & 0.194  & 0.0936 & 0.0474  & -4.21e-3 & 1.23e-4  \\
		$\Ibar$  & $\Qbar$ & 1.35   & 0.697  & -0.143  & 9.94e-2  & -1.24e-2 \\
		$\dMbar$ & $\L$   & -1.619 & 0.255  & -0.0195 & -1.08e-4 & 1.81e-5  \\
		$\dMbar$ & $\Ibar$ & -10.51 & 12.46  & -5.985  & 1.283    & -0.1044  \\
		$\dMbar$ & $\Qbar$ & -1.784 & 1.502  & -0.888  & 0.2645   & -3.55e-2 \\
		\hline\hline
	\end{tabular*}
	\caption{Numerical values of the polynomial coefficients in~\eqref{polyx}
		for the relations \relIbarL, \relQbarL and \relIbarQbar from~\cite{yagi_yunes_iloveq},
		\reldMbarL from~\cite{reina:dM},
		and \reldMbarIbar and \reldMbarQbar from~\cite{Aranguren:dM}.}
	\label{tab:polyx}
\end{table*}

\begin{table*}[htb!]
	\setlength{\tabcolsep}{15pt}
	\renewcommand{\arraystretch}{1.5}
	\centering
	\begin{tabular*}{\linewidth}{@{\extracolsep{\fill}} c S[table-format=1.2e1] S[table-format=1.2e1] S[table-format=1.2e1] S[table-format=1.2e1] S[table-format=1.2e1] }
		\hline\hline
		& \multicolumn{1}{c}{$j=0$} & \multicolumn{1}{c}{$j=1$} & \multicolumn{1}{c}{$j=2$} & \multicolumn{1}{c}{$j=3$} & \multicolumn{1}{c}{$j=4$} \\
		\hline
		$\fILrotT_{0j}$ & 1.47        & 0.0817      & 0.0149        & 0.000287      & -0.0000364  \\
		$\fILrotT_{1j}$ & 0.000167926 & 0.000190799 &	0.000114377	& -0.0000171308	& 5.46725e-7  \\
		$\fILrotT_{2j}$ & -0.57412	  & -0.178267	& -0.00436915	& 0.00247869	& -0.000088993  \\
		$\fILrotT_{3j}$ & 0.00718328  & 0.00778345  &	0.00525129	& -0.000775867	& 0.0000246931  \\
		$\fILrotT_{4j}$ & 0.0392183   & 0.0262575   &	0.00100209	& -0.000415109	& 0.0000148144  \\
		\hline
		$\fQLrotT_{0j}$ & 0.194         & 0.0936        & 0.0474        & -0.00421    & 0.000123  \\
		$\fQLrotT_{1j}$ & -0.0000559754 & -0.0000635998 & -0.0000381256 & 5.71027e-6  & -1.82242e-7  \\
		$\fQLrotT_{2j}$ & 0.191373	    & 0.0594225	    & 0.00145638	& -0.00082623 & 0.0000296643  \\
		$\fQLrotT_{3j}$ & -0.00239443	& -0.00259448	& -0.00175043	& 0.000258622 & -8.23102e-6  \\
		$\fQLrotT_{4j}$ & -0.0130728	& -0.00875252	& -0.000334029  & 0.00013837  & -4.93814e-6  \\
		\hline
		$\fIQrotT_{0j}$ & 1.35         & 0.697     & -0.143      & 0.0994     & -0.0124  \\
		$\fIQrotT_{1j}$ & 0.0000604258 & 0.0116586 & -0.00814795 & 0.00170782 & -0.000108683  \\
		$\fIQrotT_{2j}$ & -0.610433	   & -1.1022   & 0.172209	 & 0.0574922  & -0.00937998  \\
		$\fIQrotT_{3j}$ & 0.0497111	   & 0.424991  & -0.312188   & 0.0656699  & -0.0041125  \\
		$\fIQrotT_{4j}$ & 0.0629077	   & 0.152402  & 0.0378674   & -0.0414471 & 0.00533518  \\
		\hline
		$\fIQT_{0j}$ & 1.35          & 0.697     & -0.143      & 0.0994     & -0.0124 \\
		$\fIQT_{1j}$ & -0.0000483502 & 0.001354  & 0.000937207	& -0.000639553	& 0.0000788027  \\
		$\fIQT_{2j}$ & -0.481351	 & -0.922729 & 0.18146	    & 0.0409173	    & -0.00773743  \\
		$\fIQT_{3j}$ & -0.00225164	 & 0.0571651 & 0.0442787	& -0.0292549	& 0.00358643  \\
		$\fIQT_{4j}$ & 0.0251438     & 0.137069  & -0.0228351	& -0.00823763	& 0.00143439  \\
		\hline\hline
	\end{tabular*}
	\caption{The coefficients $\fILrotT_{ij}$, $\fQLrotT_{ij}$, $\fIQrotT_{ij}$, $\fIQT_{ij}$ from~\eqref{lnIbarrotTL}-\eqref{lnIbarrotQbarT}.}
	\label{tab:truncated}
\end{table*}

\begin{table*}[htb!]
	\setlength{\tabcolsep}{2.1pt}
	\renewcommand{\arraystretch}{1.5}
	\centering
	\begin{tabular*}{\linewidth}{@{\extracolsep{\fill}} c S[table-format=1.2e1] S[table-format=1.2e1] S[table-format=1.2e1] S[table-format=1.2e1] S[table-format=1.2e1] }
		\hline\hline
		& \multicolumn{1}{c}{$j=0$} & \multicolumn{1}{c}{$j=1$} & \multicolumn{1}{c}{$j=2$} & \multicolumn{1}{c}{$j=3$} & \multicolumn{1}{c}{$j=4$} \\
		\hline
		$\fdMLT_{0j}$ & -1.619& +0.255& -0.0195& -0.000108& 0.0000181\\
		$\fdMLT_{1j}$ & -0.00016792614535405588& -0.0001907993486209056& -0.0001143768468454137& 0.000017130814582213867& -5.467250623473905e-7\\
		$\fdMLT_{2j}$ & 0.574119520187566& 0.1782673776930153& 0.004369152800154278& -0.00247869053097434& 0.00008899295626715475\\
		$\fdMLT_{3j}$ & -0.007183283051656229& -0.0077834547715048665& -0.005251291543269239& 0.000775866516186344& -0.000024693063914447742\\
		$\fdMLT_{4j}$ & -0.03921834835123831& -0.02625754607210151& -0.0010020864168856143& 0.00041510858826426637& -0.000014814431123590479\\
		\hline
		$\fdMIT_{0j}$ & -10.505188729437915& +12.459951853408848& -5.985428386511496& +1.2831837436296711& -0.10443057008429853\\
		$\fdMIT_{1j}$ & -0.0022831465763007816& 0.010952400566742472& -0.010662218098550754& 0.00333842172483301& -0.0003362707394416724\\
		$\fdMIT_{2j}$ & -5.231606189031021& 7.22310840251843& -2.8743183050085994& 0.504250036605954& -0.03466558718565077\\
		$\fdMIT_{3j}$ & -0.14008488772531183& 0.5499333455868114& -0.5102177980243753& 0.15734547340009195& -0.01574927370890652\\
		$\fdMIT_{4j}$ & 0.8108330131870376& -1.0385845366850306& 0.39908457034691536& -0.06659599067275095& 0.004317315964482311\\
		\hline
		$\fdMIrotT_{0j}$ & -10.505188729437915& +12.459951853408848& -5.985428386511496& +1.2831837436296711& -0.10443057008429853\\
		$\fdMIrotT_{1j}$ & 2.133609939815832& -2.7987438547878285&  1.3942329727156606& -0.3260104013249441& 0.0312411910484838\\
		$\fdMIrotT_{2j}$ & -24.040200521846664& 38.09555479661781& -21.39205667721581& 5.511288403711999& -0.5603099649524732\\
		$\fdMIrotT_{3j}$ & 5.163957493462112& 6.397188726233331& -6.333452701904683& 0.6718937117038356& 0.1783854000809089\\
		$\fdMIrotT_{4j}$ & 8.50645072748826& -9.870498716166843& -3.3031062830327476& 4.799738701278154& -0.9735787146496931\\
		\hline
		$\fdMQT_{0j}$ & -1.7841121830792588& +1.502202541579123& -0.8879350059732071& +0.26450554816685744& -0.03553959309460048\\
		$\fdMQT_{1j}$ & -0.00017898001451295377& -0.00047058941287057984& -0.0020125294446923193& 0.0011296555639194762& -0.0001518645208255279\\
		$\fdMQT_{2j}$ & 0.4656486922391645& 0.9870544516399042& -0.26477662838667093& -0.00007650935521743328& 0.0011426823836502005\\
		$\fdMQT_{3j}$ & -0.008267666525982162& -0.01631978051712645& -0.09394305987588573& 0.0518563514009823& -0.006948987308778343\\
		$\fdMQT_{4j}$ & -0.023795797106091532& -0.14239747967575797& 0.029427238045762274& 0.0051964286916274145& -0.00097983246059603\\
		\hline
		$\fdMQrotT_{0j}$ & -1.7841121830792588& +1.502202541579123& -0.8879350059732071& +0.26450554816685744& -0.03553959309460048\\
		$\fdMQrotT_{1j}$ & -0.00028107071275083336& -0.005968072383419567& 0.0009868258675205543& 0.0015914312187976412& -0.0004165882592892514\\
		$\fdMQrotT_{2j}$ & 0.200581922325192& 1.0236191547667621& -0.049957190089184186& -0.13069469828697816& 0.025001941404074347\\
		$\fdMQrotT_{3j}$ & -0.03714109356555557& -0.23375541076861642& 0.04876420246355321& 0.06052947067370175& -0.01651037810793488\\
		$\fdMQrotT_{4j}$ & -0.07444453511466027& -0.06078704665738804& -0.04903432128321324& 0.02299035843878504& -0.000751023101315578\\
		\hline\hline
	\end{tabular*}
	\caption{The coefficients $\fdMLT_{ij}$, $\fdMIT_{ij}$, $\fdMIrotT_{ij}$, $\fdMQT_{ij}$, $\fdMQrotT_{ij}$ from~\eqref{lndMbarrotLT}-\eqref{lndMbarrotQbarrotT}.}
	\label{tab:truncated_dM}
\end{table*}

\newcommand*{\mlineleft}[1]{%
	\begingroup
	\renewcommand*{\arraystretch}{1}%
	\begin{tabular}[c]{@{}>{\raggedright\arraybackslash}l@{}}#1\end{tabular}%
	\endgroup}
\newcommand*{\mlinecenter}[1]{%
	\begingroup
	\renewcommand*{\arraystretch}{1.5}%
	\begin{tabular}[c]{@{}>{\raggedright\arraybackslash}c@{}}#1\end{tabular}%
	\endgroup}
\newcommand*{\mlineright}[1]{%
	\begingroup
	\renewcommand*{\arraystretch}{1}%
	\begin{tabular}[c]{@{}>{\raggedright\arraybackslash}l@{}}#1\end{tabular}%
	\endgroup}

\begin{table*}[htb!]
	\setlength{\tabcolsep}{2.1pt}
	\renewcommand{\arraystretch}{1.5}
	\centering
	\begin{tabular*}{\linewidth}{@{\extracolsep{\fill}} l c l}
		\hline\hline
		Term & Symbol & \multicolumn{1}{c}{Description} \\
		\hline\addlinespace[0.15cm]
		\mlineleft{Perturbative quantities} & \mlinecenter{$\chiS(\Pc,\OS)$, $\IS(\Pc)$, $\JS(\Pc,\OS)$, \\ $\MS(\Pc,\OS)$, $\QS(\Pc,\OS)$} & \mlineright{Spin parameter, moment of inertia, angular momentum, \\\quad mass, and quadrupole moment of the star computed \\\quad to second order in perturbation theory (Hartle--\\\quad Thorne model).}\\\addlinespace[0.15cm]
		\mlineleft{Fully numerical quantities} & \mlinecenter{$\chiSE(\Pc,\OS)$, $\ISE(\Pc,\OS)$, $\JSE(\Pc,\OS)$, \\ $\MSE(\Pc,\OS)$, $\QSE(\Pc,\OS)$} & \mlineright{Spin parameter, moment of inertia, angular momentum, \\\quad mass, and quadrupole moment of the star computed \\\quad to full extent, without resorting to perturbation \\\quad theory.}\\\addlinespace[0.15cm]
		\hline\addlinespace[0.15cm]
		Barred quantities & \mlinecenter{$\Ibar(\Pc)$, $\Qbar(\Pc)$} & \mlineright{Dimensionless and rotation-independent versions of \\\quad $\IS(\Pc)$ and $\QS(\Pc,\OS)$ that depend on the (non--\\\quad observable) TOV mass $\Mb(\Pc)$.}\\\addlinespace[0.15cm]
		Tilded quantities & \mlinecenter{$\Ibarrot(\Pc,\OS)$, $\Qbarrot(\Pc,\OS)$} & \mlineright{Equivalent to the barred quantities $\Ibar(\Pc)$ and $\Qbar(\Pc)$,\\\quad
			but with the change $\Mb(\Pc)\to\MS(\Pc,\OS)$. Since \\\quad $\MS(\Pc,\OS)$ is involved, these normalized quantities\\\quad  do depend on the rotation.}\\\addlinespace[0.15cm]
		\mlineleft{Tilded fully numerical \\ quantities} & \mlinecenter{$\IbarrotE(\Pc,\OS)$, $\QbarrotE(\Pc,\OS)$} & \mlineright{Equivalent to the tilded quantities $\Ibarrot(\Pc,\OS)$ and \\\quad $\Qbarrot(\Pc,\OS)$ but using the exact, fully numerical \\\quad $\JSE(\Pc,\OS)$, $\MSE(\Pc,\OS)$ and $\QSE(\Pc,\OS)$ in the \\\quad normalizations.}\\\addlinespace[0.15cm]
		\hline\addlinespace[0.15cm]
		\mlineleft{Polynomial \relIbarLQbar} & \mlinecenter{$\IbarL$, $\QbarL$} & \mlineright{Polynomial expression of the rotation-independent \\\quad \relIbarL and \relQbarL relations.}\\\addlinespace[0.15cm]
		\mlineleft{Polynomial \relIbarrotLQbarrot} & \mlinecenter{$\IbarrotL$, $\QbarrotL$, $\IbarrotQbarrot$, $\IbarrotQbar$} & \mlineright{Polynomial expression of the rotation-dependent \\\quad \relIbarrotL, \relQbarrotL, \relIbarrotQbarrot and \relIbarrotQbar \\\quad relations in terms of $\chiS$.}\\\addlinespace[0.15cm]
		\mlineleft{Polynomial \relIbarrotQbarrotE} & \mlinecenter{$\IbarrotE\rvert_{(\QbarrotE,\chiSE)}$} & \mlineright{Polynomial expression of the rotation-dependent, fully\\\quad numerical \relIbarrotQbarrotE relation.}\\\addlinespace[0.15cm]
		\hline\hline
	\end{tabular*}
	\caption{Glossary of the relevant magnitudes and symbols in the main text. The normalized counterparts of $\dM(\Pc)$, and the corresponding polynomial relations are omitted to avoid redundancy.}
	\label{tab:glossary_symbols}
\end{table*}

\begin{table*}[htb!]
	\setlength{\tabcolsep}{2.1pt}
	\renewcommand{\arraystretch}{1.5}
	\centering
	\begin{tabular*}{\linewidth}{@{\extracolsep{\fill}} l c l}
		\hline\hline
		Term & Symbol  & \multicolumn{1}{c}{Description} \\
		\hline\addlinespace[0.15cm]
		Standard approach & std & \mlineright{The coefficients $\{\ayx,\byx,\cyx,\dyx,\eyx\}$ of the original \relIbarLQbar relations, that were \\\quad computed for $\Ibar(\Pc)$ and $\Qbar(\Pc)$, are instead employed for the tilded counterparts $\Ibarrot(\Pc,\OS)$,\\\quad and $\Qbarrot(\Pc,\OS)$.}\\\addlinespace[0.15cm]
		Extended approach & ext & \mlineright{The TOV mass $\Mb(\Pc)$ that enters in the barred quantities $\Ibar(\Pc)$ and $\Qbar(\Pc)$ is computed \\\quad using the universal relations for $\dM$.}\\\addlinespace[0.15cm]
		\mlineleft{Numerical perturbative \\ approach} & NP & \mlineright{A new set of polynomial coefficients $\{\mathcal B_{ij}, \mathcal{C}_{ij}, \mathcal{D}_{ij}, \mathcal{F}_{ij}\}$ is computed for $\Ibarrot(\Pc,\OS)$ and \\\quad $\Qbarrot(\Pc,\OS)$ using the (perturbative) Hartle--Thorne model. Since the tilded quantities \\\quad depend on the rotation through $\MS(\Pc,\OS)$, the new polynomial coefficients also depend \\\quad on the rotation.}\\\addlinespace[0.15cm]
		RNS approach & RNS & \mlineright{Analogous to NP, but for the fully numerical (nonperturbative) quantities $\IbarrotE$ and $\QbarrotE$.\\\quad 
			The corresponding polynomial coefficients $\mathcal A_{ij}$ were computed in \cite{Chakrabarti:2013} using the RNS code.}\\\addlinespace[0.15cm]
		\hline\hline
	\end{tabular*}
	\caption{Glossary of all the approaches described throughout the manuscript.}
	\label{tab:glossary_approaches}
\end{table*}

\tikzstyle{start} = [rectangle, rounded corners, minimum width=3cm, minimum height=1cm, text centered, draw=blue!50!black, fill=blue!5, align=center, thick]
\tikzstyle{start_wrong} = [rectangle, rounded corners, minimum width=3cm, minimum height=1cm, text centered, draw=red!50!black, fill=red!5, align=center, thick]
\tikzstyle{decision} = [diamond, minimum width=3cm, minimum height=1cm, text centered, draw=orange!70!black, fill=orange!10, aspect=2, align=center, thin]
\tikzstyle{process} = [rectangle, minimum width=3cm, minimum height=1cm, text centered, draw=gray!80, fill=white, align=center]
\tikzstyle{arrow} = [thick,->,>=stealth, draw=gray!80!black]
\tikzstyle{dashed_line} = [dashed,thick,-, draw=gray!80!black]
\tikzstyle{ref} = [rectangle, minimum width=2cm, minimum height=0.6cm, text centered, draw=gray!80!black, dashed, fill=gray!2, font=\footnotesize, text=black, text depth=0.25ex]
\tikzstyle{ref2} = [rectangle, minimum width=3cm, minimum height=0.6cm, text centered, draw=gray!80!black, dashed, fill=gray!2, font=\footnotesize, text=black, text depth=0.25ex]

\begin{figure*}[ht]
	\centering
	\begin{tikzpicture}[node distance=2cm]
		
		\node (start) [start] {$\displaystyle\Ibar(\Pc):=\frac{\IS(\Pc)}{\Mb(\Pc)^3}$, $\displaystyle\Qbar(\Pc):=\frac{\Q(\Pc)\Mb(\Pc)}{\IS(\Pc)^2}$};
		\node (dont_know) [process, below of=start, yshift=0.4cm]{The observable quantity is $\MS(\Pc,\OS)$, not $\Mb(\Pc)$};
		\node (q1) [decision, below of=dont_know, yshift=-0.4cm] {Want to compute $\Mb(\Pc)$?};

		\node (esquina1) [left of=q1, xshift=-2cm] {};
		\node (change) [process, below of=esquina1, yshift=0.4cm] {Change $\Mb(\Pc) \to \MS(\Pc,\OS)$};
		
		\node (q2) [decision, below of=change, yshift=-0.8cm] {Keep the same coefficients \\ $\{\ayx,\byx,\cyx,\dyx,\eyx\}$?};
		
		\node (construct_rotation_dependent) [process, below of=q2, xshift=-5cm, yshift=0.4cm] {Construct new coefficients \\ using numerical Hartle--Thorne};
		\node (must_be_rotation_dependent) [process, below of=construct_rotation_dependent, yshift=0.2cm] {These new coefficients depend\\ on the rotation because \\ $\MS(\Pc,\OS)$ is involved};
		\node (rotation_dependent) [process, below of=must_be_rotation_dependent, yshift=0.2cm] {Rotation-dependent relations};
		\node (numerical) [start, below of=rotation_dependent, yshift=0.3cm] {\textbf{Numerical Perturbative}\\ \textbf{approach (NP)}};
		\node (NP_refs) [ref, below of=numerical, yshift=0.5cm] {Sec. \ref{sec:rotation}};
	
		\node (hybrid) [process, right of=must_be_rotation_dependent, xshift=3cm] {Hybrid construction:\\ rotation-dependent quantities, \\ but with the same\\ rotation-independent coefficients};
		\node (standard) [start_wrong, right of=numerical, xshift=3cm] {\textbf{Standard approach (std)}};
		\node (standard_refs) [ref2, below of=standard, yshift=0.5cm] {Secs. \ref{subsec:standard}, \ref{subsec:implications_standard}};
		
		\node (esquina2) [right of=q1, xshift=2cm] {};
		\node (universal) [process, below of=esquina2, yshift=0.4cm] {Use the relations for $\dM$ \\ to infer $\Mb(\Pc)$};
		
		\node (extended) [process, right of=q2, xshift=6cm] {\textbf{Extended approach (ext)}};
		
		\node (extended_step) [start, right of=standard, xshift=2cm] {Step-by-step \\ implementation};
		\node (extended_step_refs) [ref,below of=extended_step, yshift=0.5cm] {Sec. \ref{sec:extended}};
		
		\node (extended_rel) [start, right of=extended_step, xshift=2cm] {Rotation-dependent relations};
		\node (extended_rel_refs) [ref,below of=extended_rel, yshift=0.5cm] {Sec. \ref{rotation_dependent_extended}};
		
		\draw [arrow] (start) -- (dont_know);
		\draw [arrow] (dont_know) -- (q1);
		
		\draw [arrow] (q1) -| node[anchor=south, xshift=0.75cm] {No} (change);
		\draw [arrow] (change) -- (q2);
		\draw [arrow] (q2) -- node[anchor=west] {Yes} (hybrid);
		\draw [arrow] (q2) -| node[anchor=south, xshift=0.75cm] {No} (construct_rotation_dependent);
		\draw [arrow] (construct_rotation_dependent) -- (must_be_rotation_dependent);
		\draw [arrow] (must_be_rotation_dependent) -- (rotation_dependent);
		\draw [arrow] (rotation_dependent) -- (numerical);
		\draw [arrow] (hybrid) -- (standard);
		
		\draw [dashed_line] (numerical) -- (NP_refs);
		\draw [dashed_line] (standard) -- (standard_refs);
		
		\draw [arrow] (q1) -| node[anchor=south, xshift=-0.75cm] {Yes} (universal);
		\draw [arrow] (universal) -- (extended);
		\draw [arrow] (extended) -| (extended_step);
		\draw [arrow] (extended) -- (extended_rel);
		
		\draw [dashed_line] (extended_step) -- (extended_step_refs);
		\draw [dashed_line] (extended_rel) -- (extended_rel_refs);
		
	\end{tikzpicture}
	\caption{\textcolor{black}{Flowchart describing the different approaches to address the problem of $\Mb(\Pc)$ in the \relILQ relations.
		Since $\Mb(\Pc)$ is not an observable, one can either determine its value (right branch), or
		make the change $\Mb(\Pc)\to\MS(\Pc,\OS)$ (left branch), adapting the coefficients $\{\ayx,\byx,\cyx,\dyx,\eyx\}$ accordingly.
		The standard approach is colored in red to show that it is inaccurate (see \cite{Aranguren:dM,Aranguren:2025_proceedings}),
		as it uses $\MS(\Pc,\OS)$ while maintaining the original polynomial coefficients, that were computed for $\Mb(\Pc)$.
		The only approach not shown is RNS, which is analogous to NP, but using the fully numerical (nonperturbative) quantities \eqref{JSE}--\eqref{MSE}.}}
	\label{fig:flowchart}
\end{figure*}

\bibliographystyle{unsrt}
\bibliography{references}

\end{document}